\documentclass[12pt,a4paper]{article}
\usepackage[utf8]{inputenc}
\usepackage{fullpage}
\usepackage{cite}
\usepackage{subcaption}
\usepackage[english]{babel}
\usepackage{amsmath}
\usepackage{physics}
\usepackage{amsthm}
\usepackage{amsfonts}
\usepackage{amssymb}
\usepackage{bbold}
\usepackage{mathrsfs}
\usepackage{mathbbol}
\usepackage[toc,page]{appendix}
\usepackage{subcaption}
\usepackage{graphicx}
\usepackage{bbding}
\usepackage{esvect}
\usepackage{commath}
\usepackage{enumitem}
\usepackage[affil-it]{authblk}
\usepackage{siunitx}
\usepackage{braket}
\usepackage{phaistos}
\usepackage{multicol}
\usepackage{overpic}
\usepackage{comment}
\usepackage[colorlinks=true
  ,urlcolor=blue
  ,anchorcolor=blue
  ,citecolor=blue
  ,filecolor=blue
  ,linkcolor=blue
  ,menucolor=blue
  ,pagecolor=blue
  ,linktocpage=true
  ,pdfproducer=medialab
  ,pdfa=true
]{hyperref}
\usepackage{xcolor}

\usepackage{jheppub}
\usepackage{tikz}

\newtheorem{conjecture}{Conjecture}
\newtheorem{observation}{Observation}

\usepackage{bbm}

\newcommand{\Hil}{\mathcal{H}}
\def\ket#1{{|{#1}\rangle}} 
\newcommand{\Stab}[1]{\,\textnormal{Stab}\,#1}

\title{Entropy Cones and Entanglement Evolution for Dicke States}

\author[a]{William Munizzi,}
\author[b]{Howard J. Schnitzer}

\affiliation[a]{Department of Physics, Arizona State University,
Tempe, AZ 85281, USA}
\affiliation[b]{Martin Fisher School of Physics, Brandeis University,
Waltham, MA 02453, USA}

\emailAdd{wmunizzi@asu.edu}
\emailAdd{schnitzr@brandeis.edu}

\abstract{The $N$-qubit Dicke states $\ket{D^N_k}$, of Hamming-weight $k$, are a class of entangled states which play an important role in quantum algorithm optimization. We present a general calculation of entanglement entropy in Dicke states, which we use to describe the $\ket{D^N_k}$ entropy cone. We demonstrate that all $\ket{D^N_k}$ entropy vectors emerge symmetrized, and use this to define a min-cut protocol on star graphs which realizes $\ket{D^N_k}$ entropy vectors. We identify the stabilizer group for all $\ket{D^N_k}$, under the action of the $N$-qubit Pauli group and two-qubit Clifford group, which we use to construct $\ket{D^N_k}$ reachability graphs. We use these reachability graphs to analyze and bound evolution of $\ket{D^N_k}$ entropy vectors in Clifford circuits.}

\begin{document} 
\maketitle
\flushbottom

\section{Introduction}

There are numerous ways to classify sets of quantum states. For states in a factorizable Hilbert space $\Hil = \bigotimes^N_i \Hil_i$, one such classification is given by considering the entropy vector of each state \cite{Bao2015}. The entropy vector of a pure state $\ket{\psi} \in \Hil$ is constructed as the ordered set of all $2^N-1$ von Neumann entropies, computed by tracing out all tensor product states $\ket{\psi_i} \in \Hil_i$. While every $\ket{\psi} \in \Hil$ can be assigned an entropy vector, specifying an entropy vector does not uniquely determine a state. Instead, entropy vectors describe an equivalence relation on states in $\Hil$, assigning each state to a class based on its entanglement structure.

The space of allowed entropy vectors for a specific class of states defines the entropy cone for that class \cite{Schnitzer:2022exe,Linden2013,Bao2020,Bao2020a}. Since a particular entanglement structure often accompanies other interesting state characteristics, identifying specific entropy cones which contain the entropy vectors for different state classifications has received significant research interest. One notable case is that of holographic states, those quantum states with a smooth classical dual geometry via the AdS/CFT correspondence \cite{Maldacena:1997re,Witten:1998qj}, which possess subsystem entanglement that obeys the Ryu-Takayanagi formula \cite{Ryu:2006bv,Hubeny:2007xt}. The entropy vectors of holographic states are likewise confined to a convex polyhedral subspace of the ambient vector space, known as the holographic entropy cone. In general, having an entropy vector that is contained within a particular entropy cone is a necessary, though not sufficient, condition for a state to belong to the class described by that cone.

The $N$-qubit Dicke states are a particular class of entangled quantum state which have received notable recognition for application in quantum algorithm development and precision measurement. Dicke states have found significant use as the initial state for Quantum Approximate Optimization Algorithm (QAOA) implementation, a quantum algorithm designed to approximate combinatorial-optimization solutions \cite{farhi2014quantum}. More recently, techniques have been established to deterministically prepare $N$-qubit Dicke states using circuits of depth $\mathcal{O}(n)$ \cite{Baertschi2019}. Much of the NISQ-era utility of Dicke states for quantum computation is due to their unique entanglement properties. Certain highly-entangled Dicke states may be projected, via measurement, onto distinct lower-qubit states that cannot be locally-transformed into each other.

An alternative, and useful, classification on $\Hil$ considers how state sets transform under the action of a group. Consider a group $G \in L(\Hil)$, which transforms states $\ket{\psi} \in \Hil$. Certain elements of $G$ may act trivially on some $\ket{\psi}$, mapping the state to itself. Such elements define the stabilizer group for $\ket{\psi}$ under the action of $G$, which we denote $\Stab_G(\ket{\psi})$. The most generic quantum states are stabilized by only the identity operator $\mathbb{1}$ in a given $G$. Special sets of states, however, have larger stabilizer groups with respect to certain sets of operators. A well-known example is the set of stabilizer states, a class of classically-simulable quantum states, which are stabilized by a maximal number, specifically $2^N$, of Pauli group elements \cite{aaronson2004improved,Gottesman1997,Gottesman1998,Bravyi2004,knill2004faulttolerant,Keeler2022}.

Acting on $\ket{\psi}$ with every $g \in G$ defines the orbit $G \cdot \ket{\psi}$, which describes the trajectory of $\ket{\psi}$ through $\Hil$. For every finitely-generated $G$ the orbit is discrete, lending itself to a natural graph-theoretic description. We construct the graph which corresponds to $G \cdot \ket{\psi}$, known as a reachability graph, by assigning vertices to states in the orbit, and edges to represent generators of $G$. Viewed through the lens of quantum computation, each path in a reachability graph defines a quantum circuit and the graph itself encodes the state's evolution under all possible circuits composed of the generating gate set. States with isomorphic reachability graphs are congruent as they share an isomorphic stabilizer group under some chosen group action \cite{Keeler:2023xcx}.

We may also wish to track the evolution of specific state properties under the action of a group $G$. Analogous to the subgroups which stabilize a state, there exist elements of $G$ which leave a particular state property invariant \cite{Keeler:2023shl}, e.g. all local gates preserve entanglement structure. If our goal is to understand the dynamics of a chosen state parameter, we can restrict consideration to the subset of $G$ which non-trivially evolves that parameter. In a reachability graph representation, this restriction corresponds to eliminating vertices and edges from the graph, leaving only the graph paths, i.e. the quantum circuits, which modify the parameter under study. This modification on reachability graphs allows us to establish bounds on the dynamics of a chosen state property under circuits composed of a certain gate set.

In this paper we explore entanglement structure in Dicke states and the manner in which that entanglement can evolve through quantum circuits. In Section \ref{EntropyConeSection}, we leverage the symmetric structure of Dicke states to explicitly compute all entanglement entropies which arise in $N$-qubit Dicke systems. We use our calculation of subsystem entanglement to generate all possible Dicke state entropy vectors, which accordingly describe the $N$-qubit Dicke state entropy cone. We demonstrate inclusion, and exclusion, of Dicke state entropy vectors relative to other known entropy cones, and use our calculation to reproduce the entropy vectors for $W$ states found in \cite{Schnitzer:2022exe}. Additionally, we propose a min-cut model on weighted star graphs which realizes the symmetrized entropies of Dicke states.

In Section \ref{StabilizerSection}, we define the stabilizer group for all Dicke states under the action of the Pauli and Clifford groups. We use this set of stabilizers to construct reachability graphs which illustrate the orbit of Dicke states under circuits composed of Pauli and Clifford gates. As we are interested to understand the dynamics of Dicke state entropy vectors in Clifford circuits, we restrict to a subset of Clifford gates consisting of Hadamard and CNOT acting on two qubits. By analyzing reachability graphs built from these gates, we are able to observe bounds on entropy vector evolution directly from the reachability graphs themselves \cite{Keeler:2023shl}.

In forthcoming work, we apply the stabilizing operations identified in Section \ref{StabilizerSection} to construct error-correcting codes for logical Dicke states. One reason for using Dicke states in error-correcting codes is an increased resistance to information loss upon single-qubit thermalization. Since tracing out a single qubit from certain Dicke states recovers the same state, now defined on one lower qubit number, the redundancy of such states may yield comparative advantages in codes. 

We also consider the potential of highly-entangled Dicke states for magic distillation protocols. In this context, one benefit of Dicke states is found in their ease of preparation and consequential preference as initial states for computation. Another promising feature of using Dicke states for magic distillation relies on the significant amount of non-local magic which contained in these states. In both applications, and perhaps others not considered in this paper, we believe this exploration into Dicke state entanglement will prove useful.

\section{Review: Dicke States, Entropy Cones, and the Stabilizer Formalism}

We offer a short review of relevant background material used throughout this work. Comprehensive discussions exist for the different topics covered here, which we invite the curious reader to consult. Many significant papers discuss the structure and properties of Dicke states, as well as their utility for realizable quantum computation, of which we recommend \cite{PhysRev.93.99,Baertschi2019,nepomechie2023qudit,PhysRevA.80.052302,Stockton_2004,PhysRevLett.103.020503,PhysRevA.95.013845}. For additional details on entropy vectors and entropy cone construction, we suggest \cite{Bao2015,Hayden2013,Schnitzer:2022exe,Avis2023,Linden2013,Fadel2021}. Finally, the group-theoretic constructs presented in this section are discussed extensively in \cite{Keeler2022,aaronson2004improved,Gottesman1997,Gottesman1998,Veitch2013},and more formally in the text \cite{Alperin1995}.

\subsection{Dicke States}

The $N$-qubit Dicke states $\ket{D^N_k}$ compose an interesting class of states which can be efficiently prepared using a polynomial number of gates, despite having a larger-than-polynomial number $\binom{N}{k}$ of excitations \cite{PhysRev.93.99,Baertschi2019}. This property affords significant resource conservation compared to arbitrary state preparation, which relies on an application of $\mathcal{O}(2^N)$ gates. For this reason, Dicke states often find preference as initial states for quantum optimization algorithms, and have even been successfully implemented in experiment \cite{PhysRevA.95.013845,PhysRevLett.103.020503,Stockton_2004,PhysRevA.80.052302}. Furthermore, the highly-entangled structure of certain Dicke states can be used to project out non-locally transformable states upon measurement, such as the $GHZ$ and $W$ states, with very little computational overhead.

We construct each $N$-qubit Dicke state $\ket{D^N_k}$ as the equal superposition over all $N$-qubit states $\ket{b}$, where $b$ is a bit-string of fixed Hamming-weight $h(b) = k$. Explicitly,
\begin{equation}\label{DickeStateDefinition}
    \ket{D^N_k} \equiv \binom{N}{k}^{-1/2} \sum_{b \in \{0,1\}^n, \hspace{.1cm} h(b) = k} \ket{b}.
\end{equation}
Specific examples of Dicke states include,
\begin{equation}
    \begin{split}
        \ket{D^3_1} &= \frac{1}{\sqrt{3}} \left(\ket{100} + \ket{010} + \ket{001} \right),\\
        \ket{D^4_2} &= \frac{1}{\sqrt{6}} \left(\ket{1100} + \ket{1010} + \ket{1001} + \ket{0110} + \ket{0101} + \ket{0011} \right).
    \end{split}
\end{equation}

Dicke states of the form $\ket{D^N_1}$, those with Hamming-weight $k=1$, are exactly the $N$-qubit $W$ states $\ket{W_N}$, defined
\begin{equation}
    \ket{W_N} \equiv \frac{1}{\sqrt{N}} \left(\ket{100...00} + \ket{010...00} + ... + \ket{000...01} \right).
\end{equation}
Similarly, Dicke states of Hamming-weight $k=N$ are the $N$-qubit measurement basis state
\begin{equation}
    \ket{D^N_N} \equiv \ket{111...1} = \ket{1}^{\otimes N}.
\end{equation}

\subsection{Entropy Vectors and Entropy Cones}

We compute the entanglement entropy of a state $\rho_{\psi}$ as the von Neumann entropy
\begin{equation}\label{vonNeumannEntropy}
   S_{\psi} \equiv -\Tr \rho_{\psi} \ln \rho_{\psi}.
\end{equation}
When $\rho_{\psi}$ represents a pure state, i.e. when $\rho_{\psi} \equiv \ket{\psi}\bra{\psi}$, the property $\rho_{\psi}^2 = \rho_{\psi}$ yields total entropy $S_{\psi} = 0$. When information is measured in \textit{dits}, as with a state $\ket{\psi} \in \Hil^d$, the entropy in Eq.\ \eqref{vonNeumannEntropy} is computed using $\log_d$.

Even for an overall pure state, non-zero entanglement can exist when considering complementary subsystems of $\ket{\psi}$. For a state $\ket{\psi}$, we can consider an $\ell$-party subsystem which we denote $I$. The entanglement entropy between $I$ and its $(N-\ell)$-party complement system $\bar{I}$ is then computed
\begin{equation}
   S_{I} = -\Tr \rho_{I} \ln \rho_{I}.
\end{equation}
The object $\rho_{I}$ is the reduced density matrix of subsystem $I$, computed by tracing out its compliment $\bar{I}$.

For an $N$-party state $\ket{\psi}$, there are $2^N-1$ different subsystems we can consider. Computing $S_I$ for each subsystem $I$, and ordering the resulting set, defines the entropy vector $\Vec{S}$ for $\ket{\psi}$. For example, the entropy vector for some $3$-party pure state would have the form,
\begin{equation}\label{EntropyVectorExample}
    \Vec{S} = (S_A, S_B, S_O; S_{AB}, S_{AO}, S_{BC}; S_{ABO}).
\end{equation}
where we use a semicolon to distinguish entropies for regions of different sizes $|I|$. The final party is often labeled with $O$, as it acts a purifier for the remainder of the system.

If the overall state $\ket{\psi}$ is pure, we have the additional constraint $S_{I} = S_{\bar{I}}$, which comes from the fact that $S_{\psi} = 0$. This condition allows the entropy vector for $\ket{\psi}$ to be expressed using only $2^{N-1}-1$ entropies. Accordingly, the vector in Eq.\ \eqref{EntropyVectorExample} can be described in the reduced form
\begin{equation}\label{EntropyVectorExample2}
    \Vec{S} = (S_A, S_B; S_{AB}).
\end{equation}
We use this reduced entropy vector presentation throughout Section \ref{StabilizerSection}.

Subsystem entropies $S_I$ for multi-partite quantum states are required to obey certain entropy inequalities \cite{Bao2015,Pippenger,Hayden2013}, which can also be used to classify that state. For example, all quantum states are subadditive, meaning $S_{I} + S_{J} \geq S_{IJ}$ for all disjoint subsystems $I$ and $J$. Other entropy inequalities are more strict, and are not necessarily satisfied by generic quantum states, e.g. the monogamy of mutual information (MMI) \cite{Hayden2013} which states
\begin{equation}\label{MMIInequality}
    S_{IJ} + S_{IK} + S_{JK} \geq S_I + S_J + S_K + S_{IJK}, 
\end{equation}
for disjoint subsystems $I,\,J,$ and $K$. The MMI inequality is satisfied by all holographic states, states with a smooth classical geometric dual through the AdS/CFT correspondence.

A linear entropy inequality, such as that in Eq.\ \eqref{MMIInequality}, defines a hyperplane in some $2^N-1$ dimensional entropy-vector space, bisecting the space and placing entropy vectors which satisfy the inequality on one side, and those which fail the inequality on the other. Entropy vectors which saturate an inequality reside on the hyperplane itself. The set of linear inequalities satisfied by a class of quantum states, bounds a convex polyhedral cone in the entropy vector space, known as an entropy cone \cite{Bao2015}. Entropy vectors which correspond to a certain class of quantum states, e.g. holographic states or stabilizer states, must have an entropy vector which lies in the convex hull%
\footnote{We again highlight that this condition is necessary, but not sufficient, for identifying states corresponding to a particular class.} %
of the corresponding entropy cone. Alternatively, entropy cones can be specified by identifying all extremal rays, the entropy vectors which saturate two inequalities and lie at the intersection of two hyperplanes, or by directly identifying all possible entropy vectors for the class of states, as performed in Section \ref{EntropyConeSection}.

Entropy cones for various classes of states are well-understood for low party number. However as system size increases, so too does the number of necessary inequalities, and complexity of each inequality, needed to characterize each entropy cone. To navigate this complexity increase, much effort has turned towards studying more fundamental properties of entropy cones. The symmetrized entropy cone prescription \cite{Czech2021,Fadel2021} focuses the extremal properties of a cone's structure under a symmetry projection. Symmetrized entropies are defined as in Eq.\ \eqref{EntropyVectorExample}, with the addition of a normalization factor based on the cardinality of the subsystem. For subsystems $I$, we have
\begin{equation}\label{SymmetrizedEntropyVector}
    \Tilde{S}_k \equiv \left[ \binom{N+1}{k} \right]^{-1} \sum_{I \in \{I\}_k} S_I,
\end{equation}
where the sum is computed over all subsystems $I$, of and $N$-party states, with fixed cardinality $k = |I|$. As an example, computing the symmetrized entropy of all single-party subsystems for a $4$-party state, we have
\begin{equation}\label{SymmetrizedEntropyVector}
    \Tilde{S}_1 = \frac{1}{4}\left( S_A + S_B + S_C + S_O\right).
\end{equation}

Mathematical graphs also offer a useful description of entanglement in multi-partite quantum systems, particularly with regards to holographic systems \cite{Bao2015}. The entropy vectors of holographic states can be realized as a min-cut protocol on weighted undirected graphs, where edge cuts in the graph correspond to traversing minimal-length geodesics in the dual geometry. Broader classes of states require more generic graph descriptions, including hypergraphs \cite{Bao2020a,Bao2020} or topological links \cite{Bao2022a}. Symmetrized entropy vectors can likewise be realized using a min-cut prescription on weighted star graphs \cite{Czech2021,Fadel2021}. In Section \ref{StarGraphSection}, we extend this star graph proposal to describe the structure of Dicke state entropy vectors.

\subsection{Stabilizer Formalism and Reachability Graphs}

An essential set of gates in quantum computing is the set of Pauli gates, defined in a unitary matrix representation as
\begin{equation}\label{PauliMatrices}
    \mathbb{1}\equiv \begin{bmatrix}1&0\\0&1\end{bmatrix}, \,\, 
    \sigma_X\equiv \begin{bmatrix}0&1\\1&0\end{bmatrix}, \,\,
    \sigma_Y\equiv \begin{bmatrix}0&-i\\i&0\end{bmatrix}, \,\,
    \sigma_Z\equiv \begin{bmatrix}1&0\\0&-1\end{bmatrix}.
\end{equation}
In a fixed measurement basis $\{\ket{0},\,\ket{1}\}$, the matrices in Eq.\ \eqref{PauliMatrices} act as operators on a Hilbert space $\mathbb{C}^2$. The set $\{\sigma_X,\,\sigma_Y,\,\sigma_Z\}$ generates the $16$-element Pauli group under multiplication, denoted $\Pi_1$.

We can extend the matrix representation of Pauli gates to arbitrary qubit number by composing sets of Pauli strings. Each Pauli string describes a set of local actions performed on specified qubits in an $N$-qubit system. Every $N$-qubit Pauli string can be defined%
\footnote{This Pauli string representation as the $N$-fold tensor product of $2 \times 2$ matrices requires two conditions: first, we assume the Hilbert space factorizes into a product of $N$ qubits ($N$ copies of $\Hil^2$), and secondly, we ascribe an ordering to the set of qubits which will serve as an indexing system.} %
as a tensor product over $2 \times 2$ matrices. For example, the action of $\sigma_X$ on the $k^{th}$ qubit of an $N$-qubit system can be written
\begin{equation}\label{PauliStringExample}
    \sigma^k_X \equiv \mathbb{1}^1\otimes\ldots\otimes \mathbb{1}^{k-1} \otimes \sigma_X \otimes \mathbb{1}^{k+1} \otimes \ldots \otimes \mathbb{1}^N.
\end{equation}
Eq.\ \eqref{PauliStringExample} is an example of a weight-$1$ Pauli string, where the weight of a string denotes the number of non-identity operations in the tensor product. The $N$-qubit Pauli group $\Pi_N$ is generated by the set of all weight-$1$ Pauli strings.

Having constructed $\Pi_N$, we could further consider operations which map $\Pi_N$ to itself. The $N$-qubit Clifford group $\mathcal{C}_N$ is the set of unitaries which normalizes the Pauli group, i.e. $\mathcal{C}_N$ maps elements of $\Pi_N$ to elements of $\Pi_N$ via conjugation. In the single-qubit case, we can define $\mathcal{C}_1$ as the group generated by the Hadamard \cite{sylvester1867lx,hadamard1893resolution} and phase gates, defined as the matrices
\begin{equation}
    H\equiv \frac{1}{\sqrt{2}}\begin{bmatrix}1&1\\1&-1\end{bmatrix}, \quad P\equiv \begin{bmatrix}1&0\\0&i\end{bmatrix}.
\end{equation}

Just as with Pauli gates we can generalize to an $N$-qubit description by composing strings of Clifford operators, where we use a subscript to indicate the qubit being acted on, e.g.
\begin{equation}\label{HadamardString}
    H_k \equiv \mathbb{1}^1\otimes\ldots\otimes \mathbb{1}^{k-1} \otimes H \otimes \mathbb{1}^{k+1} \otimes \ldots \otimes \mathbb{1}^N.
\end{equation}

Unlike the Pauli group, however, the group $\mathcal{C}_N$ is not generated by only weight-$1$ Clifford strings. For $N>1$, constructing $\mathcal{C}_N$ requires the addition of the bi-local $CNOT$ gate, defined
\begin{equation}
    C_{i,j} = \begin{bmatrix}
            1 & 0 & 0 & 0\\
            0 & 1 & 0 & 0\\
	    0 & 0 & 0 & 1\\
	    0 & 0 & 1 & 0
            \end{bmatrix}.
\end{equation}
The $CNOT$ gate acts on two qubits in an $N$-qubit system by first evaluating the state of the $i^{th}$ qubit, the control bit, then performing a $NOT$ operation on the $j^{th}$ qubit, the target bit, if the $i^{th}$ qubit is found in the state $\ket{1}$. It is important to note that $C_{i,j} \neq C_{j,i}$. We may now define the group $\mathcal{C}_N$ as
\begin{equation}
    \mathcal{C}_N \equiv \langle H_1,\,...,\,H_N,\,P_1,\,...,\,P_N,\,C_{1,2},\,C_{2,1},\,...,\,C_{N-1,N},\,C_{N,N-1} \rangle.
\end{equation}

Given a Hilbert space $\Hil$ and group $G \subset L(\Hil)$, an element of $G$ is said to stabilize $\ket{\psi} \in \Hil$ if it acts trivially on $\ket{\psi}$. The set of all $g \in G$ that stabilize $\ket{\psi}$, defined
\begin{equation}
    \Stab_{G}(\ket{\psi}) \equiv \{g\in G\: | \:g\ket{\psi}=\ket{\psi}\},
\end{equation}
makes up the stabilizer subgroup of $\ket{\psi}$ under the action of $G$. Otherwise stated, $\ket{\psi}$ is a $+1$ eigenvector of each $g \in \Stab_{G}(\ket{\psi})$.

For $G$ a finite group, Lagrange's theorem \cite{Alperin1995} ensures a partition of $|G|$ for any subgroup $H \leq G$, explicitly
\begin{equation}\label{LagrangeTheorem}
    |G| = \left[G:H\right] \cdot |H|,
\end{equation}
with $[G:H]$ the index of $H$ in $G$. Furthermore when $G$ acts on a set $X$, the Orbit-Stabilizer theorem \cite{Alperin1995} gives the orbit of $x \in X$ under the action of $G$ as
\begin{equation}\label{OrbitStabilizerTheorem}
    |G \cdot x| = \left[G:\Stab_G(x)\right] = \frac{|G|}{|\Stab_G(x)|}.
\end{equation}
When $G$ acts on a Hilbert space $\Hil$, we can use Eqs. \eqref{LagrangeTheorem} and \eqref{OrbitStabilizerTheorem} to construct orbits of $\ket{\psi} \in \Hil$ under the group action \cite{Keeler:2023xcx}.

When considering the action of $\Pi_N$ on $\Hil$, states which are stabilized by a $2^N$-element subgroup of $\Pi_N$ are known as stabilizer states \cite{aaronson2004improved,Gottesman1997,garcia2017geometry,Keeler2022}. Stabilizer states play a critical role in near-term realization of quantum computing, as they represent the set of quantum systems which can be efficiently classically simulated \cite{Gottesman1998}. One way to construct the set of $N$-qubit stabilizer states is by acting on a state in the measurement basis $\{\ket{0},\ket{1}\}^N$ with the Clifford group $\mathcal{C}_N$. The set of $N$-qubit stabilizer states $\mathcal{S}_N$ is exactly the orbit of each state in $\{\ket{0},\ket{1}\}^N$, under the action of $\mathcal{C}_N$. The number of $N$-qubit stabilizer states generated by the orbit $\mathcal{S}_N$ is derived in \cite{doi:10.1063/1.4818950}, and has order
\begin{equation}\label{StabilizerSetSize}
    \left|\mathcal{S}_N\right| = 2^n \prod_{k=0}^{n-1} (2^{n-k}+1).
\end{equation}

\begin{figure}[h]
        \centering
        \includegraphics[width=10cm]{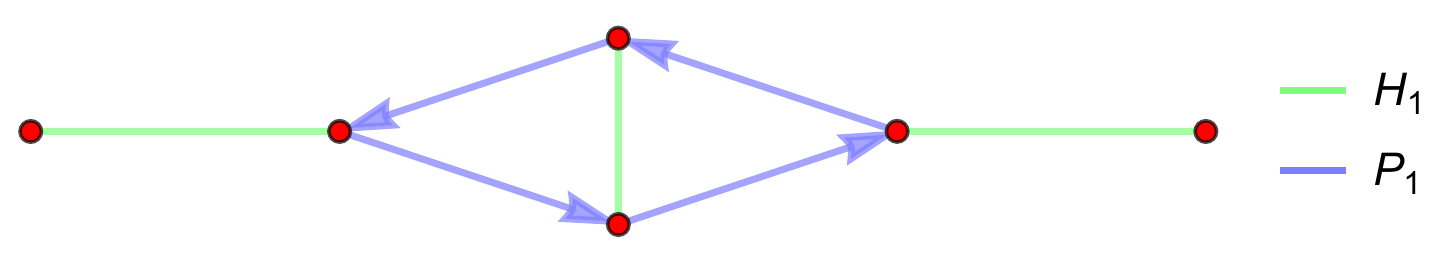}
        \caption{Orbit of $\ket{0}$ under the action of $\mathcal{C}_1$, depicted as a reachability graph. Graph vertices represent the $6$ single-qubit stabilizer states, and edges correspond to $\mathcal{C}_1$ generators. For generators which are self-inverse, e.g. the Hadamard gate, we use undirected edges.}
    \label{FullC1Graph}
    \end{figure}

The process of constructing state orbits under group action naturally admits a graph-theoretic description \cite{aaronson2004improved,Keeler:2023xcx}. For some $G$ acting on $\ket{\psi} \in \Hil$, we can assign a vertex to each state in the orbit $[G \cdot \ket{\psi}]$, and an edge to each generator of $G$. This graph is known as the reachability graph for $\ket{\psi}$, and maps the evolution of $\ket{\psi}$ through $\Hil$ under the action of $G$. Figure \ref{FullC1Graph} depicts the reachability graph for $\ket{0}$ under the single-qubit Clifford group $\mathcal{C}_1$, with vertices representing the $6$ single-qubit stabilizer states.

It is often useful to consider only the action of a subgroup $H \leq G$ on $\Hil$. Focusing on the action of $H$ highlights specific features of a state's orbit, and can better-exhibit the evolution of certain state properties through the orbit. When composing the orbit of a state $\ket{\psi}$ under the action of some $H \leq G$, the term restricted graph is sometimes used to discuss the emergent reachability graph \cite{Keeler2022}.

\section{The Entropy Cone for $\ket{D^N_k}$}\label{EntropyConeSection}

In this section, we describe the entropy cone for $N$-qubit Dicke states by explicitly building all $\ket{D^N_k}$ entropy vectors for qubit number $N$ and Hamming-weight $k$. We highlight the symmetric properties of $\ket{D^N_k}$ entropy vectors, and note the relative containment of the $\ket{D^N_k}$ entropy cone in other known entropy cones. We demonstrate that our construction for Dicke states reproduces the $W$ state entropy cone found in \cite{Schnitzer:2022exe}. Additionally, we give a realization of $\ket{D^N_k}$ entropy vectors as a min-cut prescription on weighted star graphs. In later sections we analyze the evolution of the $\ket{D^N_k}$ entropy vectors defined here, under the action of Clifford circuits.

\subsection{Dicke State Entropy Vectors}

The symmetric structure of Dicke states $\ket{D^N_k}$ enables a direct calculation of subsystem entanglement entropy from the non-zero diagonal elements of the density matrix \cite{Witten:2018zva}. For an $N$-party pure state $\ket{D^N_k}$ of Hamming-weight $k$, the entanglement entropy of an $\ell$-party subsystem is computed
\begin{equation}\label{DickeStateEntropies}
    S_{\ell}\left(\ket{D^N_k} \right) \equiv - \binom{N}{k}^{-1} \sum_{i=0}^{min(\ell,k)} \binom{\ell}{i}\binom{N-\ell}{k-i}\ln\left[\binom{N}{k}^{-1}\binom{\ell}{i}\binom{N-\ell}{k-i}\right].
\end{equation}
We can directly verify that $S_{\ell} = S_{N-\ell}$ and $S_N = 0$, from Eq.\ \eqref{DickeStateEntropies}. Furthermore, we highlight the property that $S_{\ell}\left(\ket{D^N_k}\right)$ depends only on the cardinality of a chosen subsystem.

The calculation of $S_{\ell}$ admits simplifications for specific values of $\ell$ and $k$. For states $\ket{D^N_k}$ with $\ell \geq k$, Eq.\ \eqref{DickeStateEntropies} becomes
\begin{equation}\label{lIndependentExpression}
        S_{\ell} = \ln\left[\binom{N}{k} \right] -\binom{N}{k}^{-1}\sum_{i=0}^{k}\binom{\ell}{i}\binom{N-\ell}{k-i}\ln\left[\binom{\ell}{i}\binom{N-\ell}{k-i}\right],
\end{equation}
where we note the $\ell$-independence of the first term. A derivation of \eqref{lIndependentExpression} is given in Appendix \ref{lIndependentSimplification}. A similar decoupling exists for $\ell < k$, and is shown in Eq.\ \eqref{lLessThankSimplification}.

For $k=1$, states $\ket{D^N_k}$ are the subset of $N$-qubit $W$ states, $\ket{W_N} \equiv \ket{D^N_1}$. For an $\ell$-party subsystem of $\ket{D^N_1}$, the expression for $S_{\ell}$ in Eq.\ \eqref{DickeStateEntropies} gives
\begin{equation}\label{WStateEntropies}
    S_{\ell} \left(\ket{D^N_1} \right) = \frac{\ell}{N} \ln\left[\frac{N}{\ell}\right] + \frac{(N-\ell)}{N} \ln \left[\frac{N}{N-\ell}\right],
\end{equation}
in agreement with the calculations given in \cite{Schnitzer:2022exe}.

The ordered set of all $2^N-1$ subsystem entropies for a state $\ket{D^N_k}$, computed according to Eq.\ \eqref{DickeStateEntropies}, compose the entropy vector $\vec{S}\left(\ket{D^N_k}\right)$. Since each $S_{\ell}$ depends only on $|\ell|$, all Dicke state entropy vectors share the form
\begin{equation}\label{DickeStateEntropyVector}
    \vec{S}\left(\ket{D^N_k}\right) \equiv \Bigl(\underbrace{S_1,\,...,\,S_1}_{\binom{N}{1}};\,\underbrace{S_2,\,...,\,S_2}_{\binom{N}{2}};\,...;\,\underbrace{S_{N-1},\,...,\,S_{N-1}}_{\binom{N}{N-1}};\,0\Bigl).
\end{equation}
The entropy vectors in Eq.\ \eqref{DickeStateEntropyVector} are manifestly symmetrized \cite{Czech2021}, leaving $\vec{S}\left(\ket{D^N_k}\right)$ invariant up to exchange of subsystems of equal size $|\ell|$.

All $N$-qubit Dicke state entropy vectors can be calculated using Eqs. \eqref{DickeStateEntropies} and \eqref{DickeStateEntropyVector}. Collectively, these equations describe the $N$-qubit Dicke state entropy cone, defined for each $N$ as the convex hull of all $\vec{S}\left(\ket{D^N_k}\right)$. Since each $\vec{S}\left(\ket{D^N_k}\right)$ is symmetrized, all Dicke state entropy vectors automatically satisfy the symmetrized quantum entropy cone (SQEC) inequalities \cite{Fadel2021}, 
\begin{equation}
    -S_{\ell-1} + 2S_{\ell} -S_{\ell+1} \geq 0, \qquad \forall \,1 \leq \ell \leq \lceil N/2 \rceil,
\end{equation}
which verify symmetric instances of subadditivity and strong-subadditivity. The $N$-qubit Dicke state entropy cone is therefore contained within the SQEC for all $N$. 

The monogamy of mutual information (MMI) inequality, given in Eq.\ \eqref{MMIInequality}, defines a subset of facets which bound the holographic entropy cone. For $\ket{D^N_k}$, MMI is saturated%
\footnote{MMI is trivially saturated by entropy vectors of the unentangled Dicke states $\ket{D^N_N} = \ket{1}^{\otimes N}$.} %
when $N = 3$, and violated otherwise. Similarly, $\ket{D^N_k}$ entropy vectors violate requisite inequalities for the symmetrized holographic entropy cone (SHEC), when $N > 3$, namely
\begin{equation}
    -\ell(\ell+1)S_{\ell-1}+2(\ell-1)(\ell+1)S_{\ell}-\ell(\ell-1)S_{\ell+1} \geq 0, \qquad \forall \, \ell \in [2,n/2].
\end{equation}
Consequently, portions of the $N$-qubit Dicke state entropy cone lie outside the holographic and symmetrized holographic entropy cones.

The entropy cone of stabilizer states is completely characterized up through $4$ parties ($N=5$ qubits including the purifier). It is also known that states $\ket{D^N_k}$ are not stabilizer states%
\footnote{We again note the exception for $\ket{D^N_N}$ which is trivially a stabilizer state.} %
for all $N \geq 3$. Nevertheless, we observe the following for Dicke state systems of $N \leq 5$ qubits:
\begin{observation}\label{DickeStateVectorContainment}
    The Dicke state entropy cone, for $N \leq 5$, is completely contained within the convex hull of the stabilizer entropy cone.
\end{observation}
Extending Observation \ref{DickeStateVectorContainment} to a general conjecture for all $N$ would require further knowledge of higher-party stabilizer entropy cones.

Acting with Clifford circuits on states $\ket{D^N_k}$ generates additional entropy vectors beyond those given in Eq.\ \eqref{DickeStateEntropyVector}. These Clifford group orbits of Dicke states are discussed in Section \ref{StabilizerSection}, as are the resulting entropy vectors reached under corresponding Clifford circuits. Here we note the following observation for entropy vectors generated by $2$-qubit Clifford action on $\ket{D^N_k}$:
\begin{observation}\label{CliffordOrbitDickeStateInclusion}
    All entropy vectors generated by $2$-qubit Clifford action on $\ket{D^N_k}$, for $N \leq 5$, are contained in the convex hull of the stabilizer entropy cone.
\end{observation}
While we expect Observation \ref{CliffordOrbitDickeStateInclusion} to hold for all $\ket{D^N_k}$, as well as for arbitrary Clifford circuits, we do not make an attempt towards a conjecture in this work.

We have given an explicit calculation of all $\ell$-party entanglement entropies in Dicke states $\ket{D^N_k}$, for arbitrary system size $N$ and Hamming-weight $k$. We used this result to construct all Dicke state entropy vectors $\vec{S}\left(\ket{D^N_k}\right)$, and showed that our results reproduce previous entropy vector calculations for $W$ states for $k=1$. We present the set of all $\vec{S}\left(\ket{D^N_k}\right)$, for a fixed qubit number $N$, as the $N$-qubit Dicke state entropy cone. We have highlighted that, since $\ket{D^N_k}$ entropy vectors emerge symmetrized, the $\ket{D^N_k}$ entropy cone is contained within the SQEC. At $N \geq 4$ we observed that $\vec{S}\left(\ket{D^N_k}\right)$ violates holographic inequalities, e.g. MMI, and lies outside both the holographic and symmetrized holographic entropy cones. In the next section, we use our $\ket{D^N_k}$ entropy vector construction to define a min-cut protocol which realizes Dicke state entropy vectors using weighted star graphs.

\subsection{A Graph Model for $\ket{D^N_k}$ Entropies}\label{StarGraphSection}

We now outline a protocol which compute Dicke state entropies, as given in Eq.\ \eqref{DickeStateEntropies}, as a sum over minimum-weight edge cuts on star graphs. Initial descriptions using star graphs to represent average entropies were presented in \cite{Czech:2021rxe,Fadel2021}, and later extended to include the possibility of negative edge weights in \cite{Harper:2022sky,Schnitzer:2022exe}. We demonstrate an explicit example of this star graph construction for $\ket{D^N_1}$ entropy vectors, and describe how to recursively generalize the model for $k>1$.

To construct our representation of entanglement entropy, we consider a graph $G = (V,E)$, with vertex set $V$ partitioned into subsets of internal vertices $V_{Int.} \subseteq V$, and external vertices $V_{Ext.}\subseteq V$. For an $N$-party $\ket{\psi}$ with purifier, each disjoint subsystem $\ell$ is assigned a vertex $v_{\ell} \in V_{Ext.}$, where $|V_{Ext.}| = N+1$. The entropy $S_{\ell}$ is then computed as the total weight of a min-cut on $G$ which separates $v_\ell$ from its complement subsystem $v_{\ell^C} \subset V_{Ext.}$.

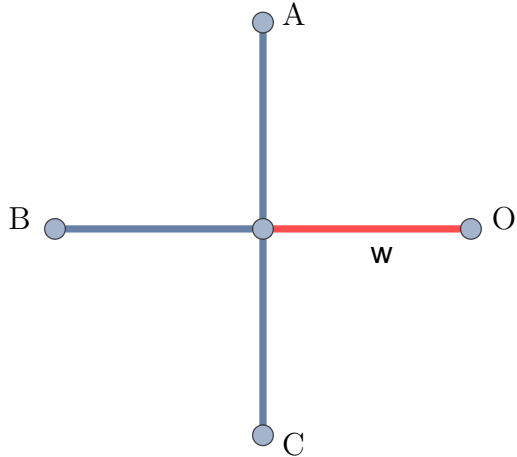
\begin{figure}[h]
		\begin{center}
		\begin{overpic}[width=6cm]{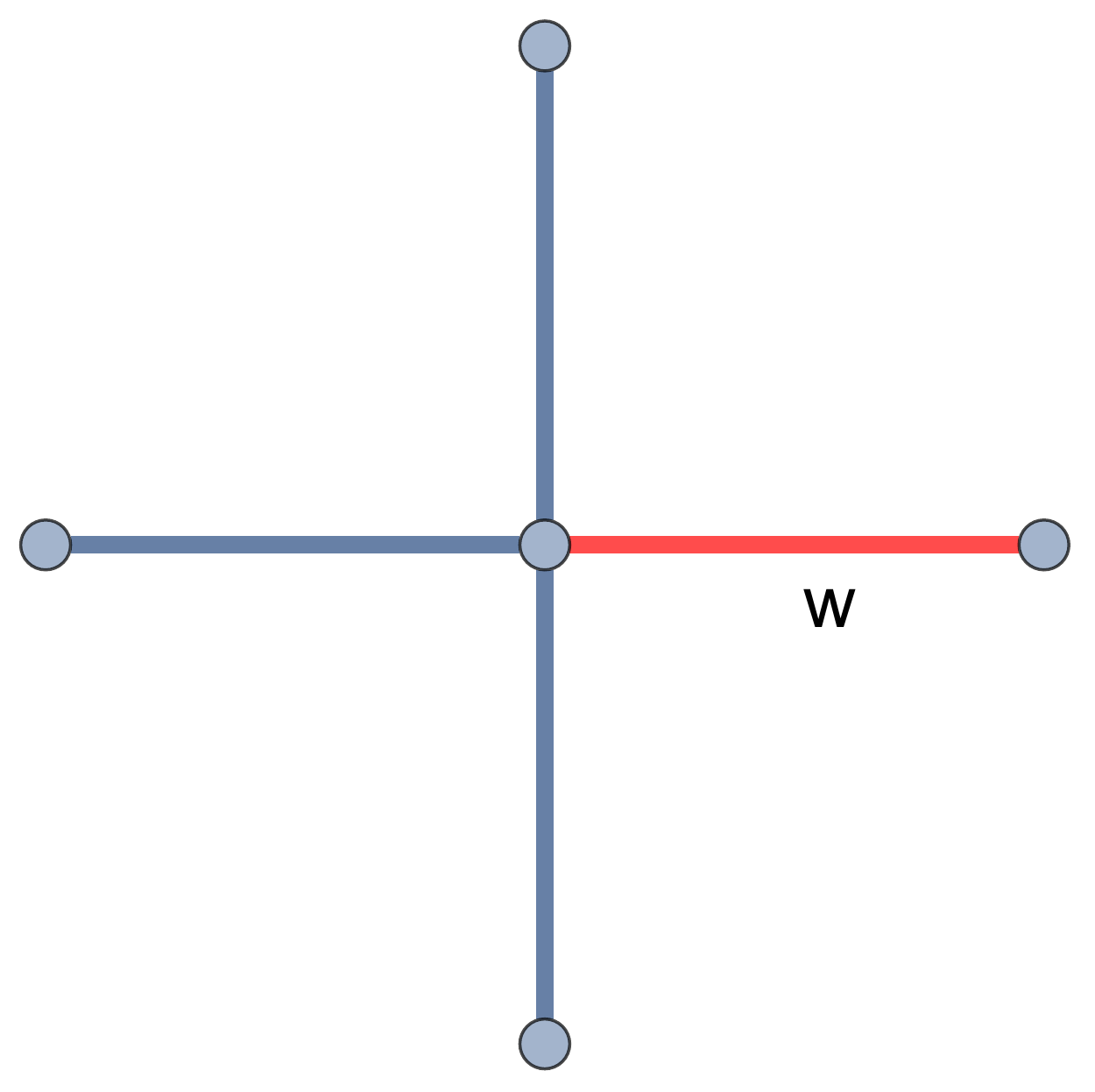}
		\put (54,95){A}
        \put (-6,50){B}
        \put (54,0){C}
        \put (100,50){O}
        \end{overpic}
        \caption{Example of a $4$-legged star graph, with $3$ legs of unit weight and one leg of weight $w \leq 0$, which realizes the entropies of $\ket{D^3_k}$. The weight $w$ is negative in this graph, and defined as a function of $k$.}
		\label{BasicStarGraph}
	\end{center}
	\end{figure}

For Dicke states $\ket{D^N_k}$ we represent each $S_{\ell}$ using a star graph with $N$ edges of unit weight, and one edge of weight $w \leq 0$. One novel feature of these graphs is that $w$ may take on select negative values, subject to the required inequalities%
\footnote{We often consider a tuple of non-negative entropies which lives in a totally non-negative sector of some $2^{N}-1$ vector space. If instead one considers a perfect tensor decomposition, negative entropies are permitted as long as they sum to a positive value in the entropy basis and satisfy the required inequalities of a chosen entropy cone. For further detail we recommend \cite{Harper:2022sky}.} %
of a particular entropy cone, which ultimately sum to non-negative entropies. Since $\ell$-party $\ket{D^N_k}$ entropies depend only on the cardinality $|\ell|$, we compute $S_{\ell}$ as the min-cut
\begin{equation}\label{SergioFadelEq}
    S_{\ell} = \min \{|\ell|,N-1 -|\ell|+w \},
\end{equation}
following \cite{Schnitzer:2022exe,Fadel2021}. Figure \ref{BasicStarGraph} gives an example of a star graph which realizes $S_{\ell}$ for a state $\ket{D^3_k}$.
The value of $w$ is defined in terms of $k$, as shown in Eqs. \eqref{StarGraphMin}--\eqref{W2Bounds}.

Since $\ket{D^N_k}$ entropies in Eq.\ \eqref{DickeStateEntropies} obey the symmetry $S_{\ell} = S_{N-\ell}$, we can define $\Tilde{S}_\ell$ to be the symmetrized variable over all $\ell$-party entanglement entropy
\begin{equation}\label{StarGraphMin}
    \Tilde{S}_\ell = \binom{N}{\ell}^{-1}\left[\binom{N-1}{\ell}S_\ell + \binom{N-1}{N-\ell}S_{N-\ell}  \right].
\end{equation}
As shown in Eq.\ \eqref{lIndependentExpression}, each $S_{\ell}$ is computed as a sum over $\ell+1$ terms when $\ell < k$, or $k+1$ terms when $\ell \geq k$. Accordingly, each $\Tilde{S}_\ell$ in Eq.\ \eqref{StarGraphMin} is realized as a sum over $\ell+1$ (or $k+1$) star graphs, for $1 \leq \ell \leq \lceil N/2 \rceil$. This sum over graphs for each $S_{\ell}$ generalizes the previous constructions in \cite{Schnitzer:2022exe} to all $\ket{D^N_k}$.

\begin{figure}[h]
        \centering
        \includegraphics[width=11cm]{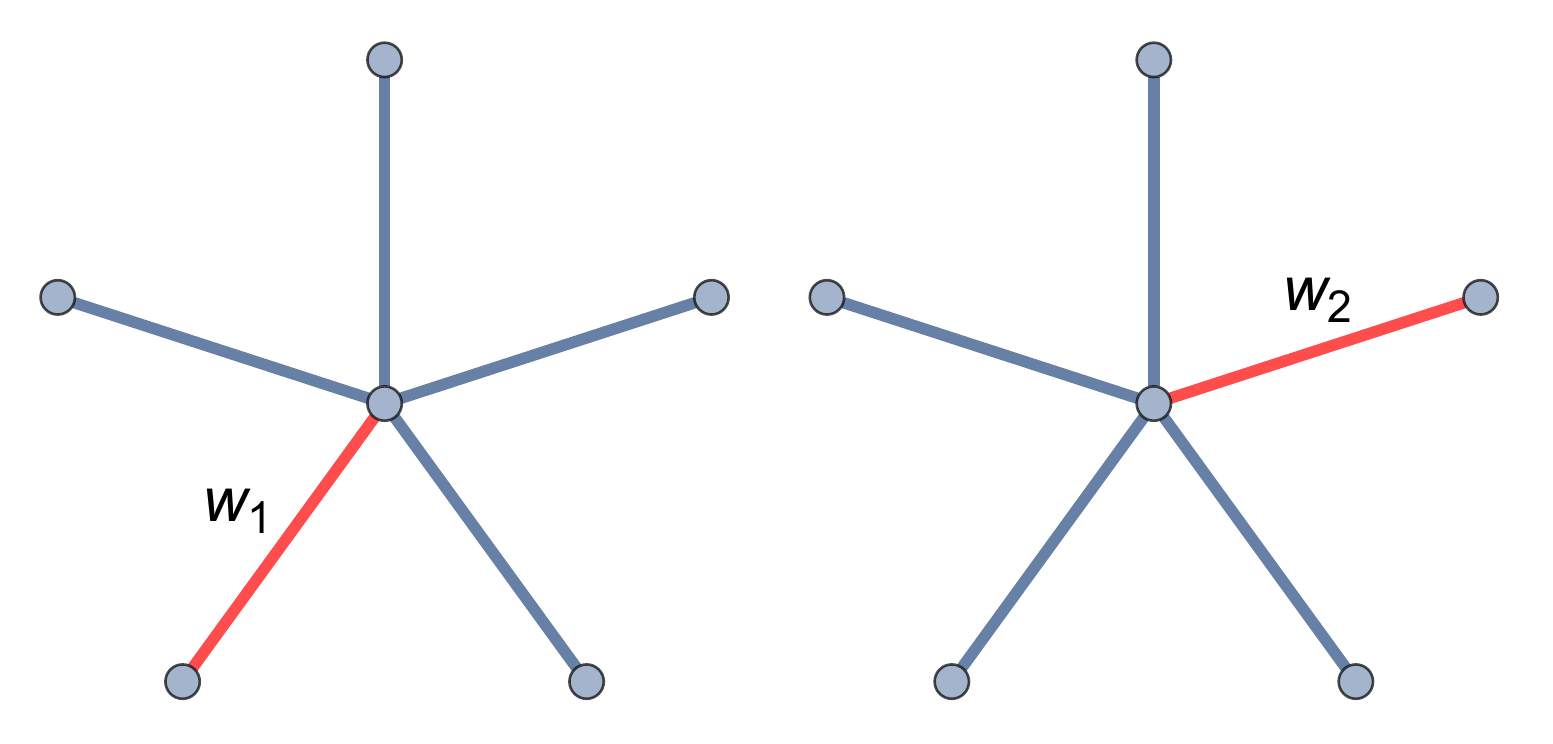}
        \caption{Pair of star graphs whose min-cut sum calculates $\Tilde{S}_{\ell}$, as in Eq.\ \eqref{WStateStarEquation}, for $\ket{D^5_1}$. The values $w_1$, $w_2$ are both negative and set by inserting Eq.\ \eqref{DickeStateEntropies} into Eq.\ \eqref{SergioFadelEq}.}
    \label{TwoStarGraphs}
    \end{figure}

To demonstrate this min-cut model, we construct an explicit representation of $S_{\ell}$ for states $\ket{D^N_1}$. Applying Eq.\ \eqref{SergioFadelEq} to Eq.\ \eqref{StarGraphMin} we have
\begin{equation}\label{WStateStarEquation}
    \Tilde{S}_{\ell} = \frac{1}{N} \left[(N-\ell)\min\{\ell,N-1-\ell+w_1\} + \ell \min \{N-\ell,w_2+\ell-1 \} \right].
\end{equation}
Figure \ref{TwoStarGraphs} shows an example pair of graphs, for the state $\ket{D^4_1}$, whose sum over min-cuts realizes Eq.\ \eqref{WStateStarEquation}. In both graphs, the negatively-weighted edge connects to the external vertex for the purifier $O$.

Beginning with the first term in Eq.\ \eqref{WStateStarEquation}, which we denote $(\Tilde{S}_{\ell})_1$, we have
\begin{equation}
    (\Tilde{S}_{\ell})_1 = \frac{1}{N}(N-\ell)\min \{ \ell,N-1-\ell+w_1\}.
\end{equation}
From Eq.\ \eqref{DickeStateEntropies} we require
\begin{equation}
    \min \{ \ell,N-1-\ell+w_1\} = \ln\left[\frac{N}{N-\ell}\right],
\end{equation}
which we solve for $w_1$ to give the bound
\begin{equation}\label{w1Solve}
    w_1 = \ell + \ln\left[\frac{N}{N-\ell}\right] - (N-1).
\end{equation}
The weight $w_1$ in Eq.\ \eqref{w1Solve} takes on negative values for
\begin{equation}\label{W1Bounds}
    \ell < (N-1) - \ln\left[\frac{N}{N-\ell}\right].
\end{equation}

Evaluating the second term $(\Tilde{S}_{\ell})_2$ in Eq.\ \eqref{StarGraphMin}, we have
\begin{equation}\label{STilde2}
    (\Tilde{S}_{\ell})_2 = \frac{\ell}{N}\min\{N-\ell,w_2+\ell-1\}.
\end{equation}
We solve Eq.\ \eqref{STilde2} to find the weight
\begin{equation}
    w_2 = \ln\left[\frac{N}{\ell}\right]-\ell+1,
\end{equation}
which is negative while
\begin{equation}\label{W2Bounds}
    \ell > 1+ \ln\left[\frac{N}{\ell} \right].
\end{equation}

The procedure in Eqs. \ref{WStateStarEquation}--\ref{W2Bounds} can be applied for all $\ket{D^N_1}$, with the resulting symmetrized entropies $\Tilde{S}_{\ell}$ described as a min-cut protocol on a pair of weighted star graphs analogous to those in Figure \ref{TwoStarGraphs}. Each graph possesses a single edge of negative weight, and the values of each weight can be determined as in Eqs. \eqref{W1Bounds} and \eqref{W2Bounds}. We now describe how to generalize this model to arbitrary $k$, by inserting sequences of star graphs to evaluate each $S_{\ell}$.

We can naturally extend the protocol described in Eqs. \ref{WStateStarEquation}--\ref{W2Bounds} to all $\ket{D^N_k}$ with $k\geq1$. For any $k >1$, each of the terms $(\Tilde{S}_\ell)_1$ and $(\Tilde{S}_\ell)_2$ in Eq.\ \eqref{StarGraphMin} are computed as a sum over min-cuts on $\ell +1$ star graphs for $\ell \geq k$, or $k+1$ star graphs for $\ell <k$. For example, consider the symmetrized entropies of $\ket{D^5_2}$,
\begin{equation}\label{5QubitEntropyRecursion}
    \begin{split}
        \Tilde{S}_1 &= (\Tilde{S}_1)_1 + (\Tilde{S}_1)_2 = 2\binom{5}{1}^{-1}\left(\frac{3}{5}\ln\left[\frac{5}{3} \right] + \frac{2}{5}\ln\left[\frac{5}{2} \right]\right) = \Tilde{S}_4,\\
        \Tilde{S}_2 &= (\Tilde{S}_2)_1 + (\Tilde{S}_2)_2 = 2\binom{5}{2}^{-1}\left(\frac{3}{5}\ln\left[\frac{5}{3} \right] + \frac{3}{10}\ln\left[\frac{10}{3} \right] + \frac{1}{10}\ln\left[\frac{10}{1} \right]\right)  = \Tilde{S}_3,\\
        \Tilde{S}_5 &= 0\\
    \end{split}
\end{equation}
The quantity $\Tilde{S}_1$ in Eq.\ \eqref{5QubitEntropyRecursion} admits a star graph representation exactly as described in Eqs. \eqref{StarGraphMin}--\eqref{W2Bounds}. Meanwhile, $\Tilde{S}_2$ is given by a sum over three star graphs, each having a single edge of negative weight.

We have given a min-cut protocol on weighted star graphs which realizes the symmetrized entropies $\Tilde{S}_{\ell}$ of all Dicke states $\ket{D^N_k}$. We gave a direct example showing graph realizations of $S_{\ell}$ for states $\ket{D^N_1}$, in agreement with results demonstrated in \cite{Schnitzer:2022exe}. We generalized this technique to $k>1$ by computing each term in  $\Tilde{S}_{\ell}$ as a sum over $\ell + 1$ star graphs, each having a single edge of negative weight. In the next section we explore group stabilizers for Dicke states under action of the Pauli and Clifford groups. We likewise analyze the orbits of Dicke states under these groups, as well as the dynamics of $\ket{D^N_k}$ entropy vectors under Clifford circuits. We illustrate $\ket{D^N_k}$ orbits as reachability graphs, using the methods given in \cite{Keeler2022,Keeler:2023xcx,Keeler:2023shl}.

\section{Stabilizers and Orbits of $\ket{D^N_k}$}\label{StabilizerSection}

In this section we construct the stabilizer subgroups for all Dicke states $\ket{D^N_k}$ under action of the Pauli and Clifford groups. We use each subgroup to construct the reachability graph for all $\ket{D^N_k}$, under the action of Pauli and Clifford group elements \cite{Keeler:2023xcx}. We highlight differences between the reachability graph structures observed for Dicke states, and those seen among the $N$-qubit stabilizer states. We later remark on the utility of the $\ket{D^N_k}$ stabilizer groups identified for error-correcting codes. We analyze $\ket{D^N_k}$ reachability graphs, with vertices colored to indicate the entropy vector, and determine the evolution of $\ket{D^N_k}$ entropy vectors under a restricted subgroup of Clifford operators. Further, we establish bounds on how much each $\ket{D^N_k}$ entropy vector can change under the select set of gates. The Mathematica data and packages used to generate all graphs is publicly available \cite{githubStab, githubCayley}.

\subsection{Pauli Group Orbits}\label{PauliStabilizers}

We first consider the action of the Pauli group%
\footnote{In the case of $\Pi_N$, as well as with $\mathcal{C}_N$, we first mod out each group by elements which act as a global phase on the group. For $\mathcal{C}_N$ this global phase element is $\omega \equiv (H_iP_i)^3$, which has the property $\omega^8 = \mathbb{1}$. Likewise for $\Pi_N$, this global phase is $\omega^2$. For details, see Section 5.1 of \cite{Keeler:2023xcx}.} %
$\Pi_N$ on the set of $N$-qubit Dicke states. While all quantum states are trivially stabilized by $\mathbb{1} \in \Pi_N$, Dicke states admit larger stabilizer subgroups in $\Pi_N$, which make them useful for stabilizer code construction. Every $\ket{D^N_k}$ is stabilized by, at least, $2$ elements of $\Pi_N$, but some are stabilized by more. 

In addition to $\mathbb{1}$, all Dicke states $\ket{D^{N}_k}$ are stabilized by the $\Pi_N$ element 
\begin{equation}\label{AllDickeStabilizer}
\begin{split}
    &\bigotimes_{i=1}^N \sigma_Z^i, \quad \textnormal{for k even},\\
    -&\bigotimes_{i=1}^N \sigma_Z^i, \quad \textnormal{for k odd}.\\
\end{split}
\end{equation}
The operator in Eq.\ \eqref{AllDickeStabilizer} acts as a $\sigma_Z$ on every qubit of an $N$-qubit system, with an additional $-1$ phase for $k$ odd. For example, Dicke states $\ket{D^{3}_1}$ and $\ket{D^{5}_2}$ have respective stabilizer subgroups given by
\begin{equation}
\begin{split}
    \Stab_{\Pi_3}(\ket{D^{3}_1}) &= \{\mathbb{1}, -\sigma_Z^1\sigma_Z^2\sigma_Z^3\},\\
    \Stab_{\Pi_5}(\ket{D^{5}_2}) &= \{\mathbb{1}, \sigma_Z^1\sigma_Z^2\sigma_Z^3\sigma_Z^4\sigma_Z^5\}.\\
\end{split}
\end{equation}

\begin{figure}[h]
        \centering
        \includegraphics[width=9cm]{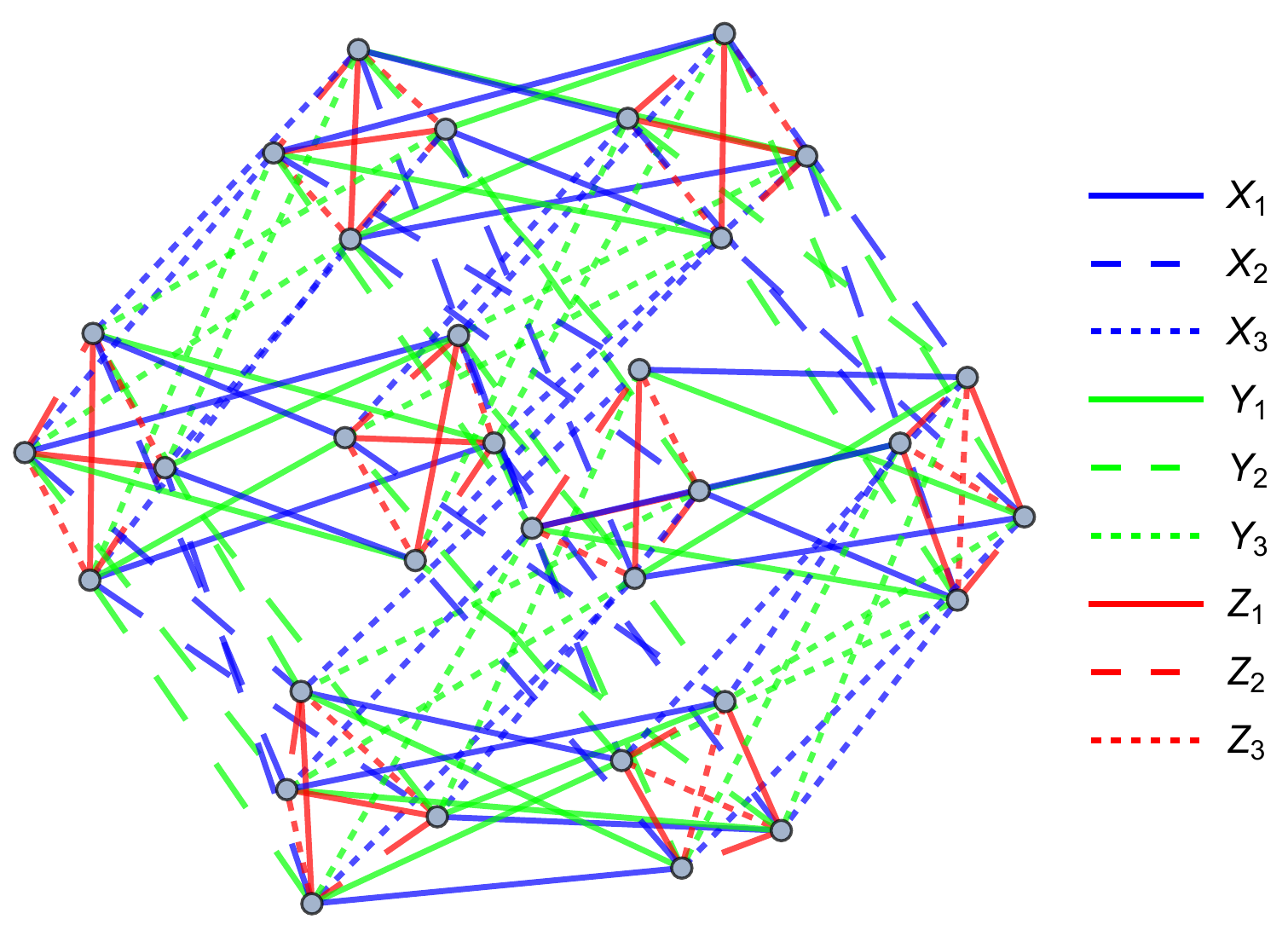}
        \caption{Orbit of $\ket{D^3_1}$ under the $3$-qubit Pauli group $\Pi_3$, which contains $32$ vertices. In general, states stabilized by only $\mathbb{1}$ and Eq.\ \eqref{AllDickeStabilizer} will have a Pauli orbit of length $2^{2n-1}$.}
    \label{D31PauliGraph}
    \end{figure}

The stabilizer subgroup containing $\mathbb{1}$ and Eq.\ \eqref{AllDickeStabilizer} quotients $\Pi_N$ into a group of order $2^{2n-1}$. Figure \ref{D31PauliGraph} illustrates the reachability graph for $\ket{D^{3}_1}$ under the action of $\Pi_3$. 

The Dicke state $\ket{D^{N}_N}$ is a stabilizer state, specifically $\ket{D^{N}_N} = \ket{1}^{\otimes N}$, as are all $\ket{D^{N}_k}$ for $N \leq 2$. Accordingly, $\ket{D^{N}_N}$ is stabilized by a $2^N$-element subgroup of $\Pi_N$. In addition to $\mathbb{1}$ and the operator in Eq.\ \eqref{AllDickeStabilizer}, $\ket{D^{N}_N}$ is stabilized by the action of $-\sigma_Z$ on any single qubit, as well as $\sigma_Z$ on any qubit pair for $N \geq 2$. Written explicitly, the set
\begin{equation}\label{StabilizerNN}
    \Stab_{\Pi_N}(\ket{D^{N}_N})\supseteq\{-\sigma_Z^i,\, \sigma_Z^i\sigma_Z^j\}, \quad \forall \, i,j \in \{1,N\}.
\end{equation}

\begin{figure}[h]
        \centering
        \includegraphics[width=9cm]{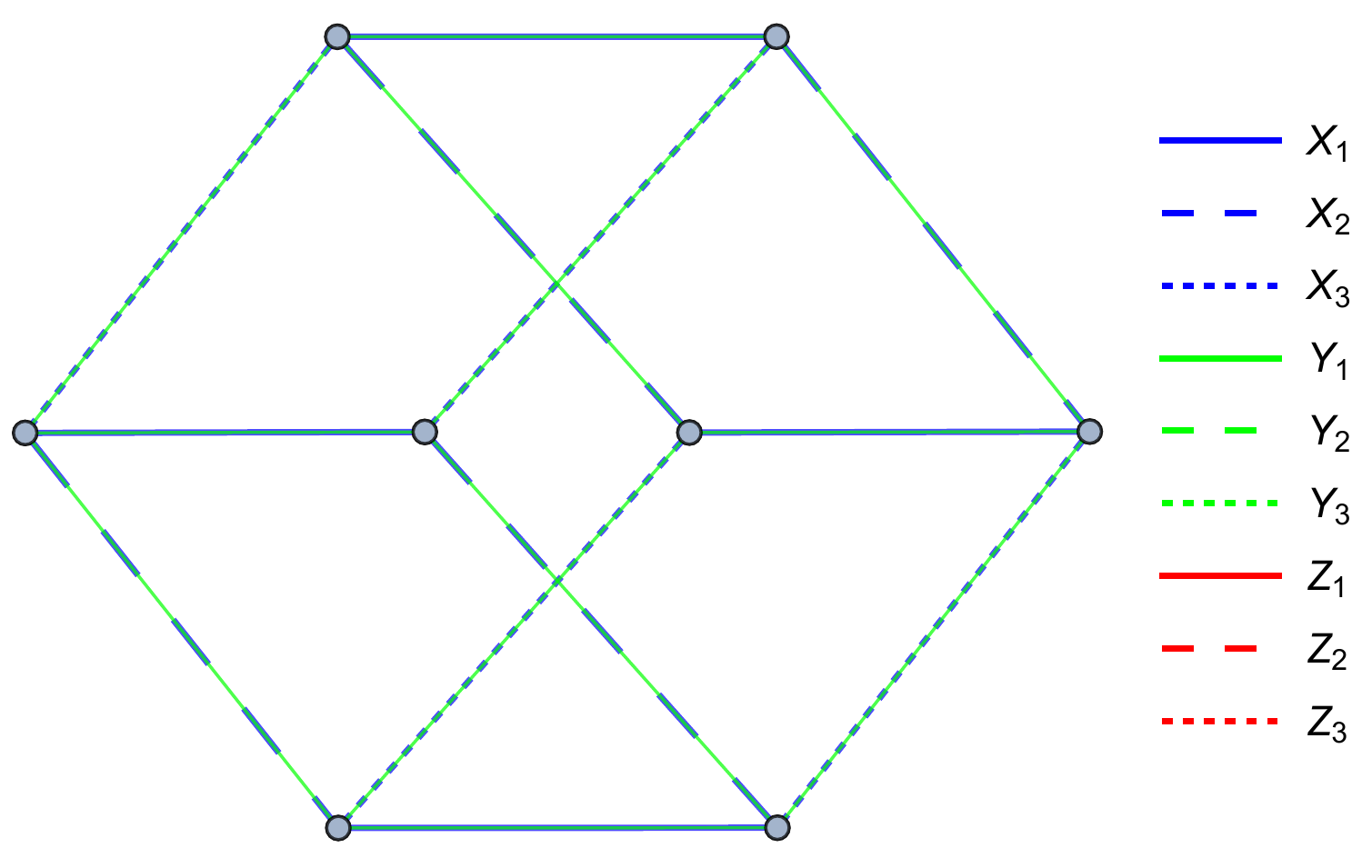}
        \caption{Reachability graph of state $\ket{D^3_3}$ under the action of $\Pi_3$. This graph contains $8$ vertices, with gates $\sigma_X^i$ and $\sigma_Y^i$ acting the same on $\ket{D^3_3}$. States of the form $\ket{D^N_N}$ are stabilizer states and are stabilized by $2^N$ elements of $\Pi_N$.}
    \label{D33PauliGraph}
    \end{figure}

As an example, the stabilizer subgroup of $\ket{D^{3}_3}$ consists of the $6$ operations
\begin{equation}
    \Stab_{\Pi_3}(\ket{D^{3}_3}) = \{\mathbb{1},\, -\sigma_Z^1,\,-\sigma_Z^2,\,-\sigma_Z^3, \,\sigma_Z^1\sigma_Z^2, \,\sigma_Z^1\sigma_Z^3,\,\sigma_Z^2\sigma_Z^3,\, -\sigma_Z^1\sigma_Z^2\sigma_Z^3\}.
\end{equation}
The corresponding Pauli orbit for $\ket{D^{3}_3}$ is shown in Figure \ref{D33PauliGraph}, where we note that all edges of the graph simultaneously represent the actions of $\sigma_X^i$ and $\sigma_Y^i$, as they act identically on $\ket{D^{3}_3}$ up to global phase.

Finally, we consider Dicke states $\ket{D^{N}_k}$ where $N=2k$. States $\ket{D^{2k}_k}$ are stabilized by the simultaneous action of $\sigma_X$ and $\sigma_Y$ on every qubit. For states $\ket{D^{2k}_k}$, we have
\begin{equation}\label{Stabilizer2kk}
\Stab_{\Pi_N}(\ket{D^{2k}_k}) \supset \left\{
    \bigotimes_{i=1}^{2k} \sigma_{X}^i,\, \bigotimes_{i=1}^{2k} \sigma_{Y}^i \right\},
\end{equation}
as well as $\mathbb{1}$ and Eq.\ \eqref{AllDickeStabilizer}. The two additional stabilizers in Eq.\ \eqref{Stabilizer2kk} result in $4$-element stabilizer subgroup for $\ket{D^{2k}_k}$ under the action of $\Pi_N$. 

For the example state $\ket{D^{4}_2}$, its stabilizer subgroup under $\Pi_4$ can be written
\begin{equation}
\Stab_{\Pi_4}(\ket{D^{4}_2}) =\{\mathbb{1},\,\sigma_X^1\sigma_X^2\sigma_X^3\sigma_X^4,\,\sigma_Y^1\sigma_Y^2\sigma_Y^3\sigma_Y^4,\,\sigma_Z^1\sigma_Z^2\sigma_Z^3\sigma_Z^4\}.
\end{equation}
Since $\ket{D^{4}_2}$ is stabilized by a $4$-element subgroup of $\Pi_4$, its orbit under $\Pi_4$, depicted in Figure \ref{D42PauliGraph} of Appendix \ref{AdditionalGraphs}, reaches $64$ states.

We have given the stabilizer subgroup for all $\ket{D^{N}_k}$ under action of the $N$-qubit Pauli group. We used the stabilizer subgroup to generate a reachability graph for $\ket{D^{N}_k}$, which represents each state's orbit under $\Pi_N$. In the following section we extend our analysis to consider action of the $N$-qubit Clifford group $\mathcal{C}_N$, as well as $\mathcal{C}_N$ subgroups. We use the reachability graphs of $\ket{D^{N}_k}$ to analyze entanglement structures observed in Dicke state orbits.

\subsection{Clifford Group Orbits and Entanglement Evolution}

Dicke states $\ket{D^N_k}$ are not stabilizer states for $N \geq 3$ and $N \neq k$. However, interestingly, states $\ket{D^N_k}$ are stabilized by more Clifford group elements than just $\mathbb{1}$. In this section, we extend our study of $\ket{D^N_k}$ orbits by considering the action of the two-qubit Clifford group $\mathcal{C}_2$. Since entanglement modification via Clifford gates occurs through bi-local action, this restriction to $\mathcal{C}_2$ is sufficient for exploring the evolution of Dicke state entropy vectors under Clifford circuits. We construct the stabilizer subgroup for each $\ket{D^N_k}$ under the action of $\mathcal{C}_2$, and compute the size of each orbit.

We also present reachability graphs for $\ket{D^N_k}$ under the action of the $\mathcal{C}_2$ subgroup $(HC)_{1,2} \equiv \langle H_1,\, H_2,\, C_{1,2},\, C_{2,1} \rangle$, with vertices colored by entropy vector as in \cite{Keeler2022,Keeler:2023xcx}. Since the $P_1$ and $P_2$ gates cannot modify a state's entropy vector, the subgroup $(HC)_{1,2}$ contains all non-trivial entropy vector dynamics. Furthermore, graph representations of $(HC)_{1,2}$ orbits are easier to parse than $\mathcal{C}_2$ orbits, as they contain a factor of $10$ less vertices. We use $(HC)_{1,2}$ orbits of Dicke states to give a bound on the number of times $\ket{D^N_k}$ entropy vectors can change under this gate set. A more general bound on entropy vector dynamics using quotient graphs is derived in \cite{Keeler:2023shl}.

All one and two-qubit Dicke states, $\ket{D^1_1}$ and $\ket{D^2_k}$, are also stabilizer states. Likewise, every $\ket{D^N_N}$ is a stabilizer state as well. Accordingly, states $\ket{D^2_k}$ and $\ket{D^N_N}$ are stabilized by a $192$-element subgroup of $\mathcal{C}_2$, and their reachability graph is exactly the two-qubit stabilizer state graph shown in Figure \ref{FullC2Graph} of Appendix \ref{AdditionalGraphs}. 

\begin{figure}[h]
		\begin{center}
		\begin{overpic}[width=13cm]{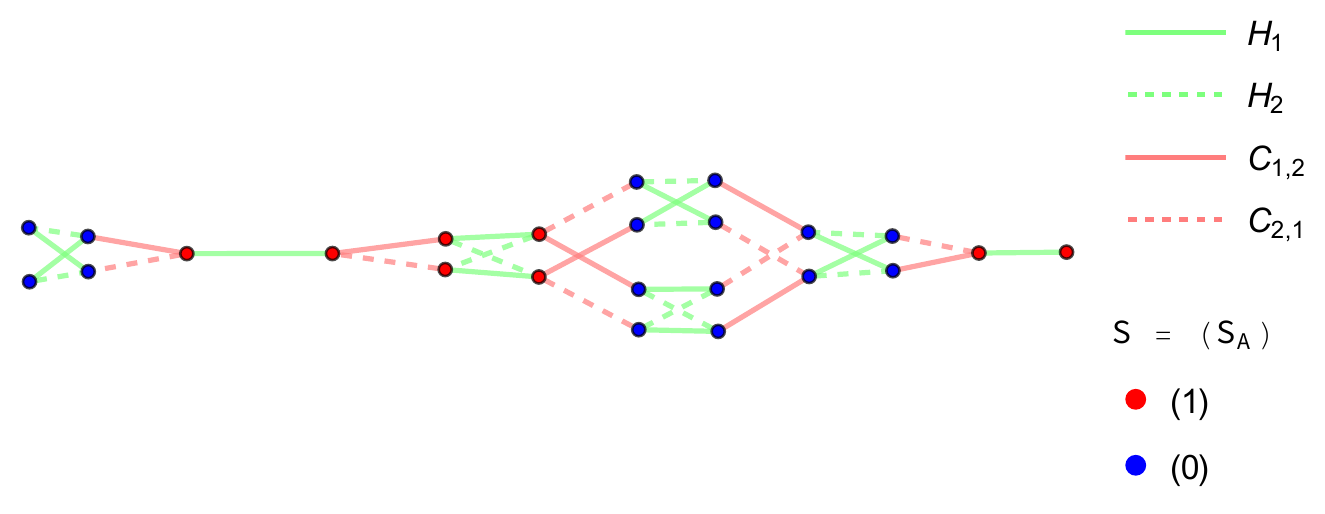}
		\put (61.5,23) {\footnotesize{$\swarrow \ket{D^N_N}$}}
        \put (13.5,16) {\footnotesize{$\uparrow \ket{GHZ}_N$}}
        \end{overpic}
        \caption{Orbit of $\ket{D^N_N}$ under $\langle H_1,\,H_2,\,C_1,\,C_2\rangle$ subgroup. This reachability graph has $24$ vertices and $2$ entanglement possibilities, unentangled and maximally-entangled. Since $\ket{D^N_N} = \ket{1}^{\otimes N}$, this reachability graph is shared by a subset of the $N$-qubit stabilizer states.}
		\label{HCGraphD21}
	\end{center}
	\end{figure}

When we restrict to the action of $(HC)_{1,2} = \langle H_1,\,H_2\,C_{1,2}\, C_{2,1}\rangle$, the orbit of $\ket{D^2_k}$ and $\ket{D^N_N}$ contains $24$ states. Figure \ref{HCGraphD21} illustrates this orbit, showing the reachability graph of $\ket{D^N_N}$ under the action of $(HC)_{1,2}$. While the state $\ket{D^N_N}$ is unentangled, elements of the $(HC)_{1,2}$ subgroup are capable of generating instances of $GHZ$-type entanglement throughout the orbit.

Dicke states of the form $\ket{D^N_1}$, which define the set of $N$-qubit $W$ states, as well as states $\ket{D^N_{N-1}}$, are stabilized by $4$ elements of $\mathcal{C}_2$. Specifically, the states $\ket{D^N_1}$ and $\ket{D^N_{N-1}}$ have stabilizer subgroup
\begin{equation}\label{DN1StabSubgroup}
\begin{split}
        \Stab_{\mathcal{C}_2}(\ket{D^N_1}) &= \{\mathbb{1},\,H_2C_{1,2}H_2,\,C_{1,2}C_{2,1}C_{1,2},\,H_2C_{1,2}H_2C_{1,2}C_{2,1}C_{1,2}\},\\
        & = \Stab_{\mathcal{C}_2}(\ket{D^N_{N-1}}).
\end{split}
\end{equation}

The stabilizer group in Eq.\ \eqref{DN1StabSubgroup} yields an orbit of $2880$ states for $\ket{D^N_1}$ and $\ket{D^N_{N-1}}$, under the action of $\mathcal{C}_2$. 

\begin{figure}[h]
		\begin{center}
		\begin{overpic}[width=13cm]{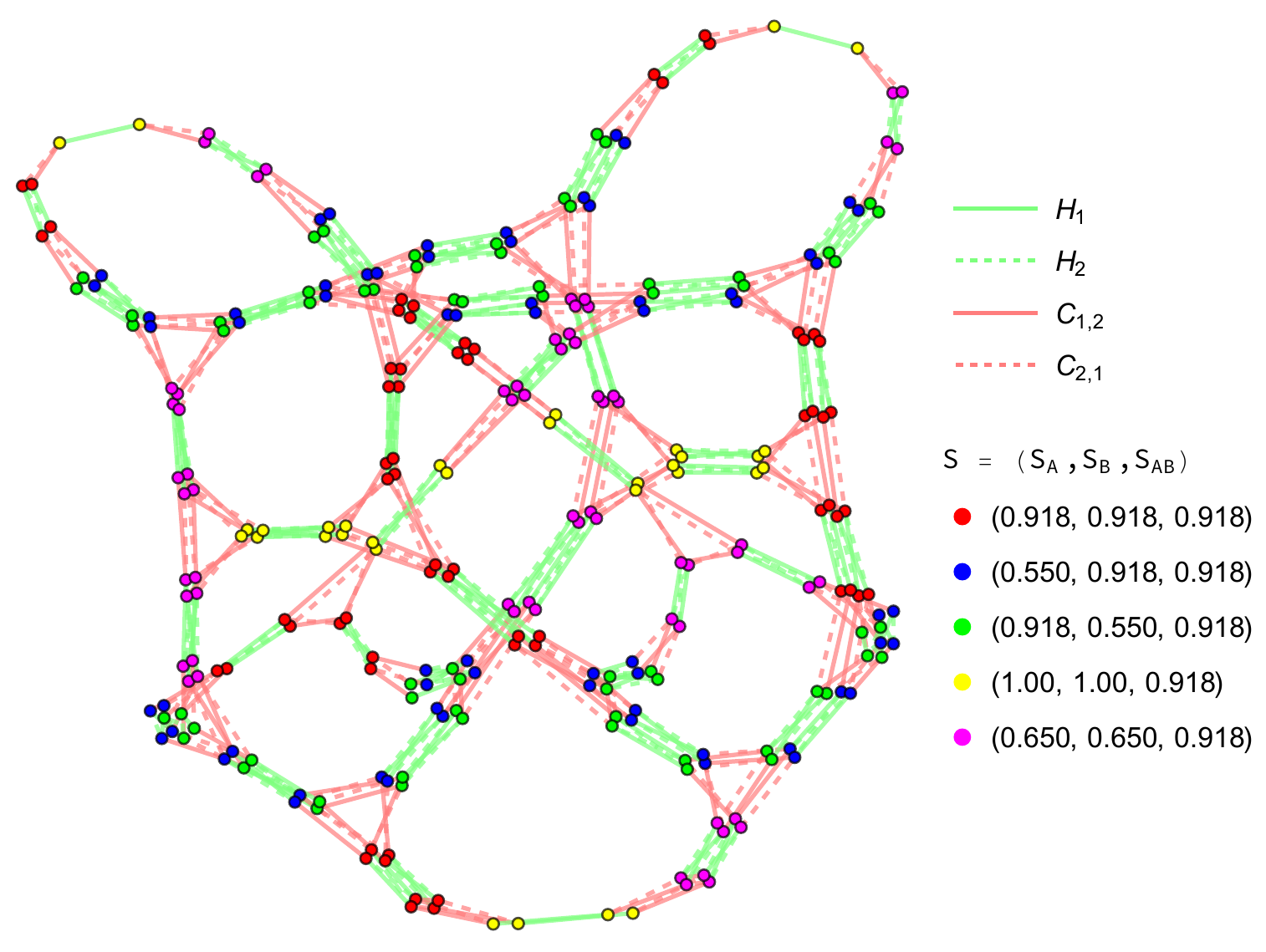}
		\put (53,67.9) {\footnotesize{$\leftarrow \ket{D^3_1}$}}
        \end{overpic}
        \caption{Orbit of $\ket{D^3_1}$ under $\langle H_1,\,H_2,\,C_1,\,C_2\rangle$ action. The graph has $288$ vertices, and contains $5$ different entropy vectors. We especially note the topological distinction of this graph, compared to the $288$-vertex stabilizer state graph. Numerical approximations for entanglement entropies are shown in the figure, with exact values in Table \ref{tab:EntropyVectorTable1}.}
		\label{HCGraphD31}
	\end{center}
	\end{figure}

Restricting group action to $(HC)_{1,2}$, the orbits of all $\ket{D^N_{1}}$ and $\ket{D^N_{N-1}}$, for $N \geq 3$, consist of $288$ states. Figure \ref{HCGraphD31} depicts the $\ket{D^N_{1}}$ reachability graph under $(HC)_{1,2}$, shown for the example state $\ket{D^3_{1}}$. While this reachability graph has $288$ vertices, it is not isomorphic to the $288$-vertex graph observed for stabilizer states under the action of $(HC)_{1,2}$, presented in \cite{Keeler2022}.

\begin{figure}[h]
        \centering
        \includegraphics[width=14cm]{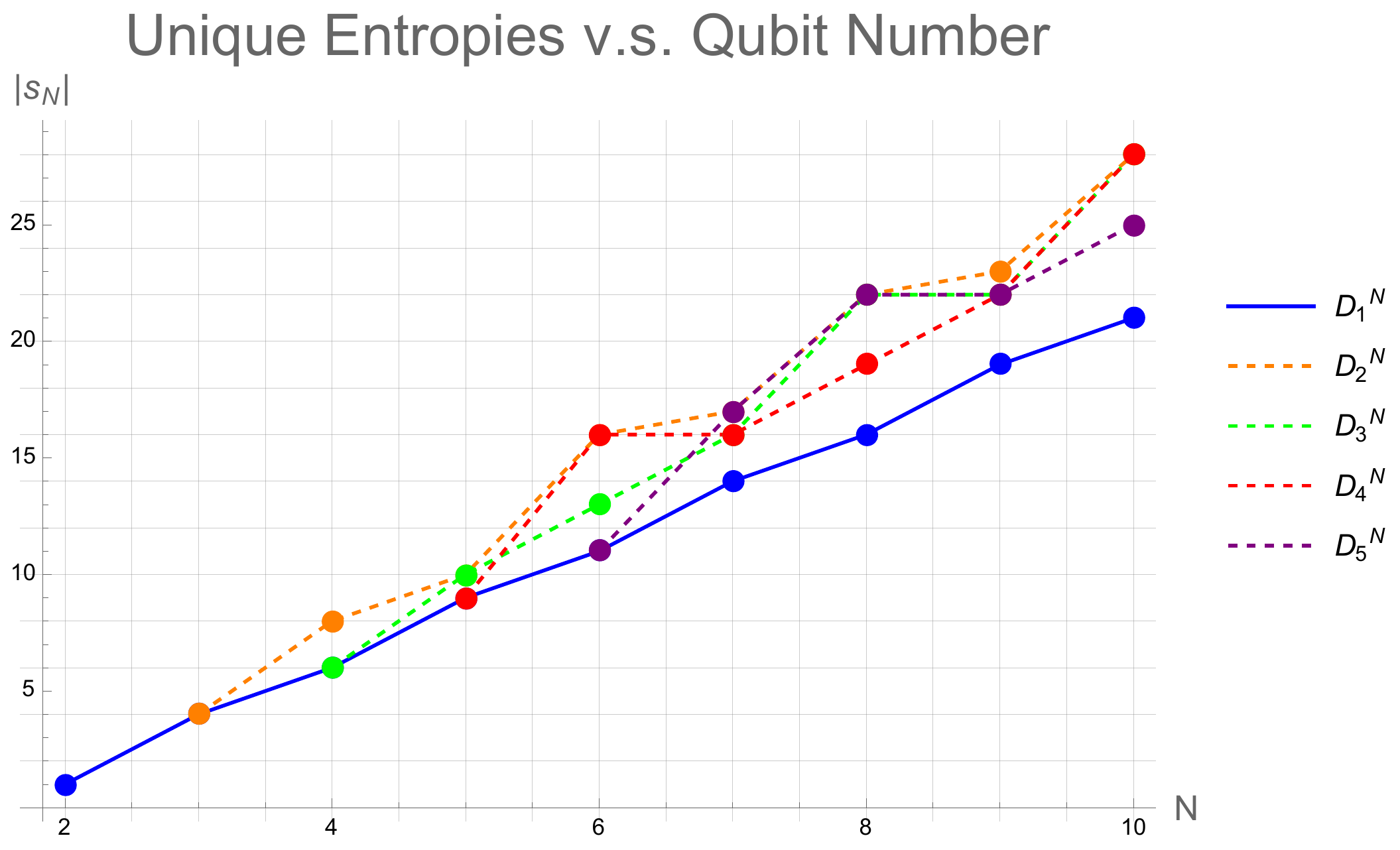}
        \caption{The number of unique entanglement entropies $|s_N|$ comprising all entropy vectors in the $(HC)_{1,2}$ orbit of $\ket{D^N_k}$. We plot this entanglement entropy cardinality against increasing qubit number $N$, for $N \leq 10$ and $1 \leq k \leq 5$. The solid blue line depicts the special case of $\ket{D^N_1}$ described in Conjecture \ref{EntanglementCardinality}.}
    \label{DickeEntropiesQubitNum}
    \end{figure}

The orbit of $\ket{D^N_{1}}$ under $(HC)_{1,2}$, for all $N \geq 3$, contains $5$ different entropy vectors. As described in \cite{Keeler:2023shl}, there are maximally $5$ unique entropy vectors that can be generated for states $\ket{D^N_{1}}$, using all $(HC)_{1,2}$ circuits in Figure \ref{HCGraphD31}. While the number of entropy vectors in graphs like Figure \ref{HCGraphD31} cannot increase beyond $5$, the number of different entanglement entropies comprising those entropy vectors, denoted $|s_N|$, continues to grow with increasing qubit number $N$. In Figure \ref{HCGraphD31}, the entropy vectors in the orbit of $\ket{D^3_1}$ are built of $4$ distinct entanglement entropies, shown to the right of the figure. For arbitrary $N$-qubit states $\ket{D^N_1}$, we conjecture the following:
\begin{conjecture}\label{EntanglementCardinality}
For $N \geq 2$, the number of unique entanglement entropies which comprise all entropy vectors in the $(HC)_{1,2}$ orbit of $\ket{D^N_1}$ increases as
\begin{equation}
    |s_N| = \lfloor \frac{5N-7}{2} \rfloor. 
\end{equation}
\end{conjecture}
The state $\ket{D^1_1}$ is pure and has zero entanglement entropy. The number of unique entanglement entropies encountered in the $(HC)_{1,2}$ orbit of $\ket{D^N_k}$ are depicted in Figure \ref{DickeEntropiesQubitNum}, for $N \leq 10$ qubits.

All remaining Dicke states $\ket{D^{N}_k}$, with $1 < k < N-1$, are stabilized by a $2$-element subgroup of $\mathcal{C}_2$. The stabilizer subgroup for such $\ket{D^{N}_k}$ states is given by
\begin{equation}\label{AllOtherDStabilizer}
\Stab_{\mathcal{C}_2}\left(\ket{D^N_k}\right) = \{\mathbb{1},\, C_{1,2}C_{2,1}C_{1,2},\}, \quad \forall \, 1 < k < N-1,
\end{equation}
Consequently, the $\mathcal{C}_2$ orbit of states stabilized by Eq.\ \eqref{AllOtherDStabilizer} reaches $5760$ states.

\begin{figure}[h]
		\begin{center}
		\begin{overpic}[width=15cm]{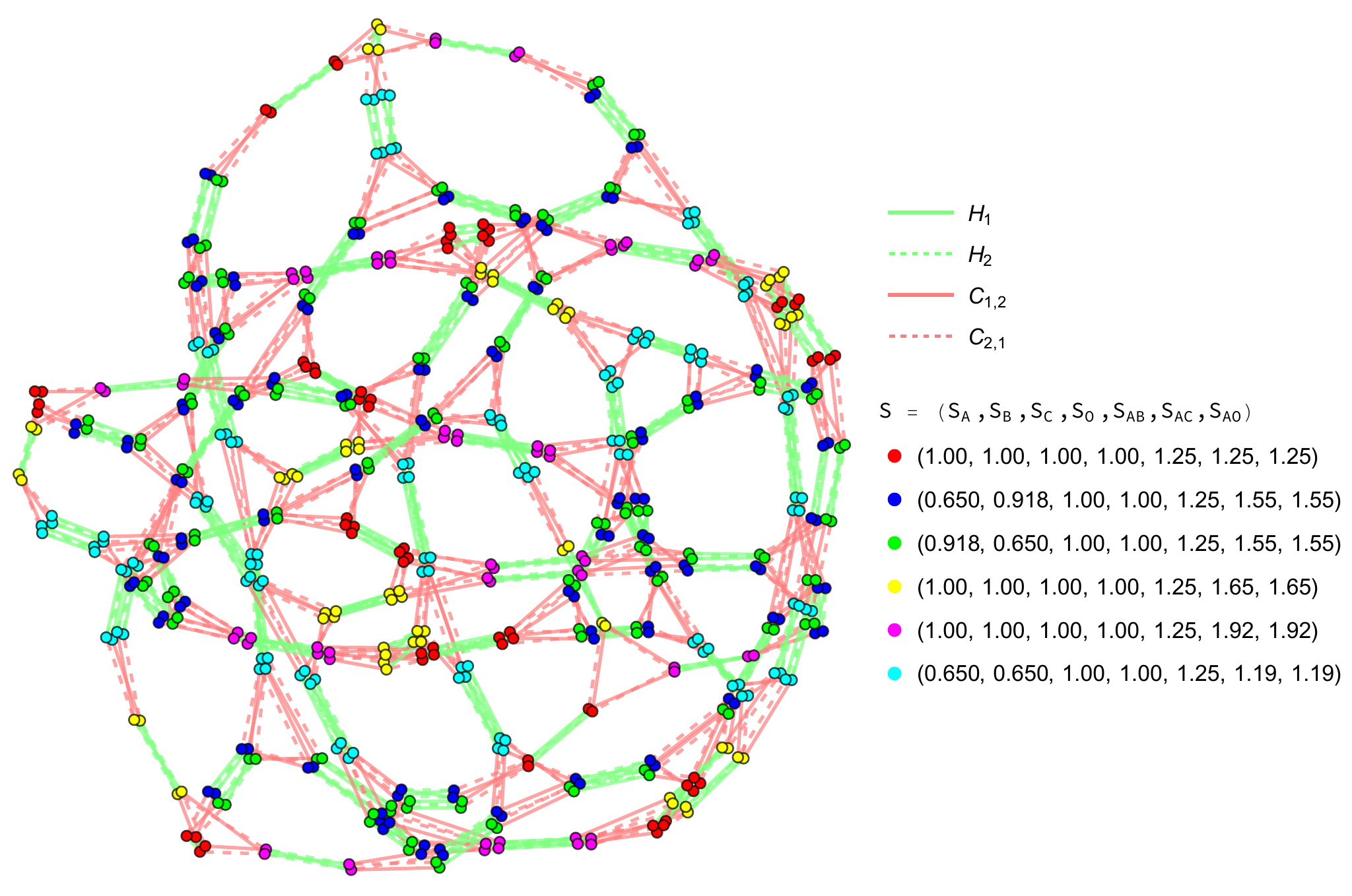}
		\put (19.6,55.5) {\footnotesize{$\uparrow \ket{D^4_2}$}}
        \end{overpic}
        \caption{Reachability graph showing the orbit of $\ket{D^4_2}$ under $(HC)_{1,2}$. This reachability graph has $576$ vertices, a vertex count never observed among stabilizer states, and contains $6$ different entropy vector possibilities. We provide numerical approximations for the entropy vector component in the figure, with exact values given in Table \ref{tab:EntropyVectorTable2}.}
		\label{HCGraphD42}
	\end{center}
	\end{figure}

The action of $(HC)_{1,2}$ on $\ket{D^{N}_k}$, for $1 < k < N-1$, generates an orbit of $576$ states. This $576$-element orbit under $(HC)_{1,2}$ is particularly interesting as it differs in size from any stabilizer state orbit under $(HC)_{1,2}$ action \cite{Keeler2022,Keeler:2023xcx}. Stated alternatively, the stabilizer subgroup in Eq.\ \eqref{AllOtherDStabilizer} is not shared by any stabilizer state at any qubit number. As a result, reachability graphs with $576$ vertices, like that in Figure \ref{HCGraphD42}, are never witnessed for stabilizer states.

For $k > 1$, the number of unique entanglement entropies that make up entropy vectors in the $(HC)_{1,2}$ orbit of $\ket{D^N_k}$ increases with system size. Figure \ref{DickeEntropiesQubitNum} illustrates the relationship between cardinality $|s_N|$ and qubit number $N$, for $\ket{D^N_k}$ up to $N=10$ qubits.

Reachability graphs like that in Figure \ref{HCGraphD42} admit $6$ unique entropy vectors throughout the orbit. The bounds proposed in \cite{Keeler:2023shl} limit graphs isomorphic to Figure \ref{HCGraphD42} to having, at most, $9$ different entropy vectors. However, since entanglement dynamics additionally depends on the state being evolved through the quantum circuit, the symmetries of Dicke states constrain the number of entropy vectors in these graphs to $6$. As with the orbits of $\ket{D^N_1}$, while the overall number of entropy vectors in the reachability graph is fixed, for all $N$ and $k$, the number of distinct entanglement entropies which make up those vectors continues to increase for larger and larger $N$, and varies for different values of $k$.

We have identified the stabilizer subgroups for all Dicke states $\ket{D^N_k}$, under the action of the $N$-qubit Pauli group $\Pi_N$, as well as the two-qubit Clifford group $\mathcal{C}_2$. We demonstrated that there exist three distinct stabilizer subgroups for states $\ket{D^N_k}$, depending on the values of $N$ and $k$. States $\ket{D^N_N} = \ket{1}^{\otimes N}$ belong to the set of $N$-qubit stabilizer states, and share the corresponding stabilizer groups \cite{Keeler2022,Keeler:2023xcx}. States $\ket{D^N_1}$ and $\ket{D^N_{N-1}}$ share a stabilizer subgroup in $\Pi_N$ and $\mathcal{C}_2$, as do all $\ket{D^N_k}$ with $1 < k < N-1$. We illustrated the orbit of $\ket{D^N_k}$, under the action of $\Pi_N$ using reachability graphs.

In order to understand the evolution of Dicke state entropy vectors, we likewise constructed reachability graphs for all $\ket{D^N_k}$ under the action of the $\mathcal{C}_2$ subgroup $(HC)_{1,2} = \langle H_1,\,H_2,\,C_1,\,C_2\rangle$. Since entanglement modification in Clifford circuits occurs through bi-local action, restricting to this subgroup enabled us to place constraints on the dynamics of $\ket{D^N_k}$ entropy vectors under Clifford gates. We found the number of entropy vectors in each $(HC)_{1,2}$ orbit to be constant, with $5$ entropy vectors possible on graphs of $288$ vertices, and $6$ entropy vectors on graphs with $576$ vertices. While the number of entropy vectors on these graphs is fixed, the number of distinct entropies continued to increase.  

\section{Discussion}

In this work we constructed the entropy cone for $N$-qubit Dicke states $\ket{D^N_k}$ by calculating all entropy vectors for arbitrary values of $N$ and $k$. We first defined a function to compute the entanglement entropy $S_{\ell}$ of any $\ell$-party subsystem of $\ket{D^N_k}$. We demonstrated that $\ket{D^N_k}$ entropy vectors are manifestly symmetric, with $S_{\ell}$ only dependent on the size of subsystem $\ell$, and therefore lie within the convex hull of the SQEC. We likewise find that $\ket{D^N_k}$ entropy vectors are contained within the Stabilizer entropy cone as far as it is characterized, up to $N=5$. Dicke state entropy vectors do not, however, satisfy the necessary HEC or SHEC conditions for all $N \geq 3$, where $k \neq N$. We verified that our calculation accurately reproduces all vectors of the $N$-qubit $W$ state entropy cone \cite{Schnitzer:2022exe}, since $\ket{W_N} = \ket{D^N_1}$. 

We additionally define a prescription which realizes average entropies $\Tilde{S}_{\ell}$, for $\ket{D^N_k}$, as a min-cut protocol on weighted star graphs. Entropies $S_{\ell}$, as in Eq.\ \ref{DickeStateEntropies}, are computed as the minimum-weight edge cut on $\ell+1$ star graphs of $N+1$ legs each. Every star graph has a single edge of weight $w \leq 0$, with the precise value of $w$ constrained by the values of $N$ and $\ell$. The sum of two set of star graphs, one representing $S_{\ell}$ and one $S_{N-\ell}$, defines $\Tilde{S}_{\ell}$ for all $\ket{D^N_k}$. This graph representation of $\ket{D^N_k}$ entropies builds upon other min-cut protocols for symmetrized entropies \cite{Czech:2021rxe,Fadel2021,Schnitzer:2022exe}, and is an interesting direction for future study.

We studied the orbits of $\ket{D^N_k}$ under action of $N$-qubit Pauli group $\Pi_N$ and the $2$-qubit Clifford group $\mathcal{C}_2$. Interestingly, Dicke states form a set of non-stabilizer states which are stabilized by more Clifford elements than just $\mathbb{1}$. We identified the stabilizer subgroup for every $\ket{D^N_k}$, under the action of both groups, which we used to generate $\ket{D^N_k}$ reachability graphs \cite{Keeler:2023xcx}. Each $\ket{D^N_k}$ reachability graph depicts the state's orbit under the action of a chosen group, 
and generates all states which can be reached through circuits built of the generating gates. Since $\ket{D^N_k}$ is often initialized as the starting state for many quantum algorithms, 
the reachability graphs in Section \ref{StabilizerSection} provide a map through the Hilbert space for algorithms that begin with $\ket{D^N_k}$. Furthermore, since construction of reachability graphs is not limited to the Clifford group \cite{Munizzi_2022}, it would be interesting to explore Dicke state orbits under a universal set of gates. 

Reachability graphs can also be used to bound entanglement evolution under a chosen set of gates, by examining how many times the entropy can be changed by circuits in the graph \cite{Keeler:2023shl}. Motivated to explore the dynamics of $\ket{D^N_k}$ entropy vectors under $\mathcal{C}_2$, we focused on the subgroup $(HC)_{1,2} = \langle H_1,\,H_2,\,C_{1,2},\,C_{2,1} \rangle$ since entanglement in Clifford circuits occurs, at most, through the bi-local CNOT gate. We found that the number of entropy vectors on each $\ket{D^N_k}$ reachability graph is constant, with $5$ entropy vectors on the $288$-vertex graphs, like that in Figure \ref{HCGraphD31}, and $6$ entropy vectors on the $576$-vertex graphs, like Figure \ref{HCGraphD42}. While the number of entropy vectors is fixed, the number of distinct entanglement structures which compose each vector continues to increase for larger and larger systems.

We expect our analysis of the $N$-qubit Dicke state entropy cone and $\ket{D^N_k}$ orbits to generalize for qudit Dicke states. Circuits for deterministically preparing arbitrary qudit Dicke states are known, and many recursive generalizations Dicke state properties have also been demonstrated. We expect the  $\ket{D^N_k}$ entropy vectors presented in this paper to generalize similarly, with preliminary efforts towards $W$ state entropy cone generalization given in \cite{Schnitzer:2022exe}. The Pauli and Clifford groups can likewise be extended to arbitrary Hilbert space dimension \cite{Jagannathan:2010sb,Hostens_2005}, which would allow us to extend our orbit model and consider the evolution of entanglement evolution for higher-dimensional Dicke systems. 

The Dicke state stabilizers presented in this work find immediate application in stabilizer code construction. Given a scheme for encoding logical Dicke states and a suitable choice of measurement, we can construct an error-correcting channel directly using the stabilizers in Section \ref{StabilizerSection}. For specific Dicke states, such as $\ket{D^N_k}$, the entanglement structure renders the state robust to single-qubit loss, particularly at large $N$. We expect this characteristic to offer significant error-correction advantages when using Dicke state encoding for noisy processing. In future work, we explore this proposal and construct a class of Dicke stabilizer codes, evaluating their performance when compared to existing schemes.

Finally, the entanglement structure of certain Dicke states makes $\ket{D^N_k}$ an interesting candidate for magic distillation protocols \cite{Bravyi2004}. Coupled with the ease of preparing $\ket{D^N_k}$, Dicke states can enable improved protocols for distilling magic with minimal overhead. The states $\ket{D^5_1}$, $\ket{D^5_2}$, and $\ket{D^5_4}$ specifically possess a significant amount non-local magic \cite{Bao2022a}, though ultimately experience error rates slightly above the fault-tolerant Bravyi-Kitaev threshold. These error rates can be improved however, by passing $\ket{D^5_k}$ through a short sequence of gates. This further motivates an understanding $\ket{D^N_k}$ orbits under universal gate sets, which can provide circuits to improve the utility of Dicke states in distillation schemes.

\section*{Acknowledgements}

HJS wishes to thank Matt Headrick for alerting him to the possible interest overlaps of the Arizona State University group with his. WM wishes to thank Adam Burchardt, ChunJun Cao, Jonathan Harper, Cynthia Keeler, and Jason Pollack for helpful discussions. WM is supported by the U.S. Department of Energy under grant number DE-SC0019470 and by the Heising-Simons Foundation ``Observational Signatures of Quantum Gravity'' collaboration grant 2021-2818.
Dicke states were brought to our attention by Rafael Nepomechie in \cite{nepomechie2023qudit}, and in private communication to HJS.

\newpage

\begin{appendices}

\section{Simplification for fixed $\ell$ and $k$}\label{lIndependentSimplification}

In this Appendix we give a few reductions for the $\ket{D^N_k}$ entropies in Eq.\ \eqref{DickeStateEntropies}. Expanding $\ln$ in Eq.\ \eqref{DickeStateEntropies} gives,
\begin{equation}
\begin{split}
    S_{\ell} = - \binom{N}{k}^{-1}\sum_{i=0}^{min(\ell,k)}& \binom{\ell}{i}\binom{N-\ell}{k-i}\ln\left[\binom{N}{k}^{-1}\right]\\
    &-  \binom{N}{k}^{-1}\sum_{i=0}^{min(\ell,k)}\binom{\ell}{i}\binom{N-\ell}{k-i}\ln\left[\binom{\ell}{i}\binom{N-\ell}{k-i}\right].\\
\end{split}
\end{equation}
We invoke the Chu-Vandermonde identity, which reads
\begin{equation}
    \sum_{i=0}^n \binom{r}{i}\binom{s}{n-i} = \binom{r+s}{n},
\end{equation}
to simplify the first sum as
\begin{equation}\label{DecoupledDickeEntropy}
\begin{split}
    S_{\ell} = \binom{N}{k}^{-1}& \binom{N}{\min(\ell,k)}\ln\left[\binom{N}{k} \right] \\
    &  -\binom{N}{k}^{-1}\sum_{i=0}^{min(\ell,k)}\binom{\ell}{i}\binom{N-\ell}{k-i}\ln\left[\binom{\ell}{i}\binom{N-\ell}{k-i}\right].\\
\end{split}
\end{equation}
For the case when $\ell \geq k$, Eq.\ \eqref{DecoupledDickeEntropy} further simplifies to,
\begin{equation}
        S_{\ell} = \ln\left[\binom{N}{k} \right] -\binom{N}{k}^{-1}\sum_{i=0}^{k}\binom{\ell}{i}\binom{N-\ell}{k-i}\ln\left[\binom{\ell}{i}\binom{N-\ell}{k-i}\right],
\end{equation}
where we note the $\ell$-independence of the first term. 

Likewise for $\ell < k$ we have the reduction
\begin{equation}\label{lLessThankSimplification}
        S_{\ell} = \frac{k!(N-k)!}{\ell!(N-\ell)!}\ln\left[\binom{N}{k} \right] -\binom{N}{k}^{-1}\sum_{i=0}^{\ell}\binom{\ell}{i}\binom{N-\ell}{k-i}\ln\left[\binom{\ell}{i}\binom{N-\ell}{k-i}\right].
\end{equation}

\newpage

\section{Additional Reachability Graphs}\label{AdditionalGraphs}

\subsection{Clifford Orbits of $\ket{D^N_N}$}

\begin{figure}[h]
        \centering
        \includegraphics[width=11cm]{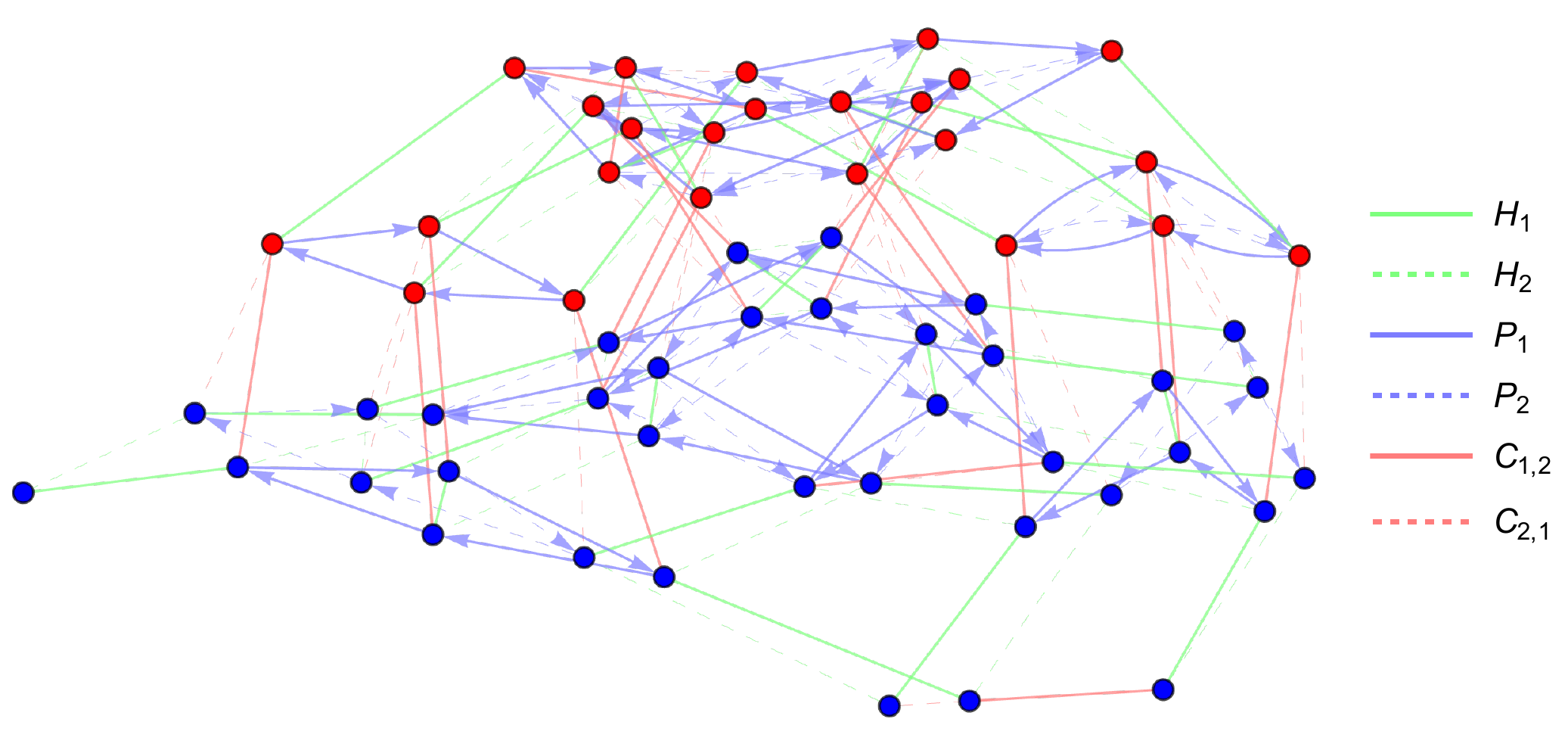}
        \caption{Orbit of all $\ket{D^N_N}$ under the action of the $2$-qubit Clifford group $\mathcal{C}_2$. The state $\ket{D^N_N}$ is a stabilizer state, and therefore its reachability graph is isomorphic to that of all $2$-qubit stabilizer states.}
    \label{FullC2Graph}
    \end{figure}

The reachability graph for $\ket{D^N_N}$, and accordingly for all stabilizer states, under the action of $\mathcal{C}_2$ is shown in Figure \ref{FullC2Graph}.

\begin{figure}[h]
        \centering
        \includegraphics[width=15cm]{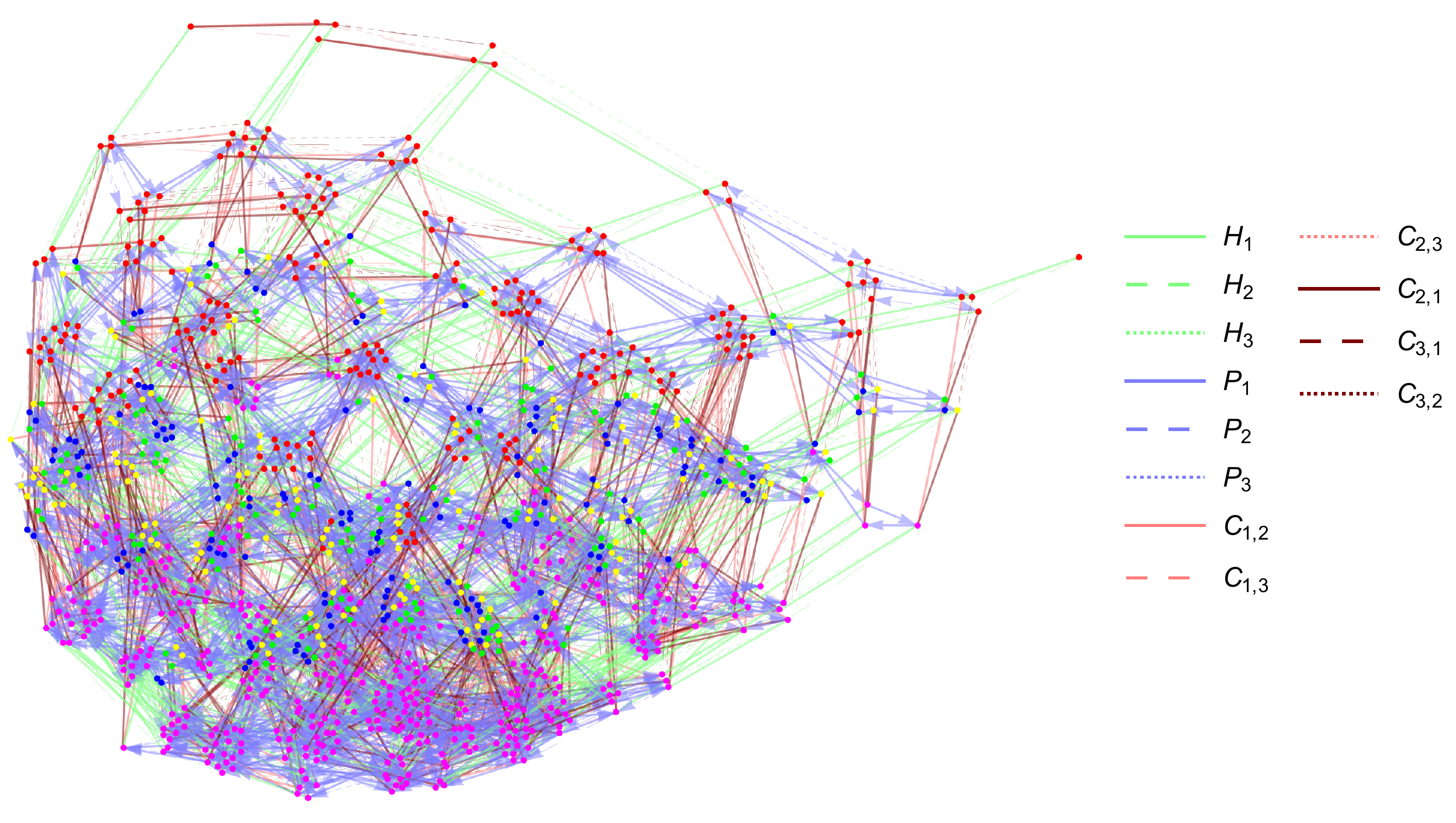}
        \caption{Orbit of $\ket{D^N_N}$ under the action of the $3$-qubit Clifford group $\mathcal{C}_3$. Since $\ket{D^N_N}$ is a stabilizer state, this reachability graph is isomorphic to the orbit shared by all $3$-qubit stabilizer states.}
    \label{FullC3Graph}
    \end{figure}

Figure \ref{FullC3Graph} gives the reachability graph for $\ket{D^N_N}$ under the action of $\mathcal{C}_3$.

\begin{figure}[h]
        \centering
        \includegraphics[width=12cm]{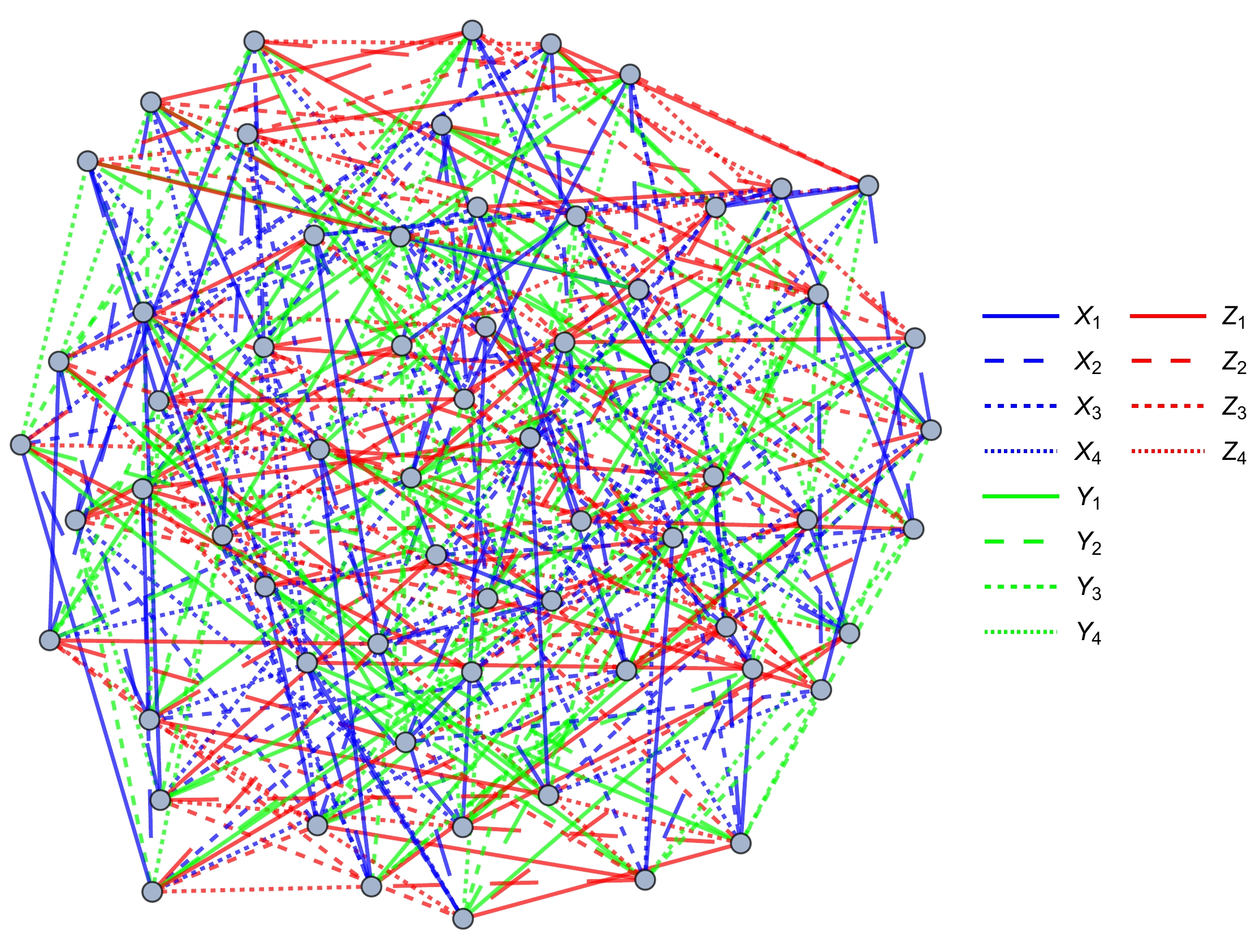}
        \caption{Orbit of $\ket{D^4_2}$ under the action of $\Pi_4$. This reachability graph contains $64$ vertices as $\ket{D^4_2}$ is only stabilized by $4$ elements of $\Pi_4$.}
    \label{D42PauliGraph}
    \end{figure}

Figure \ref{D42PauliGraph} depicts the orbit of state $\ket{D^{4}_2}$ under action of the $4$-qubit Pauli group $\Pi_4$. Since $\ket{D^{4}_2}$ is stabilized by a $4$-element subgroup of $\Pi_4$, its reachability graph contains $64$ vertices.

\subsection{$(HC)_{1,2}$ Orbits of Higher-Qubit $\ket{D^N_1}$}\label{HigherN1Appendix}

\begin{figure}[h]
        \centering
        \includegraphics[width=14cm]{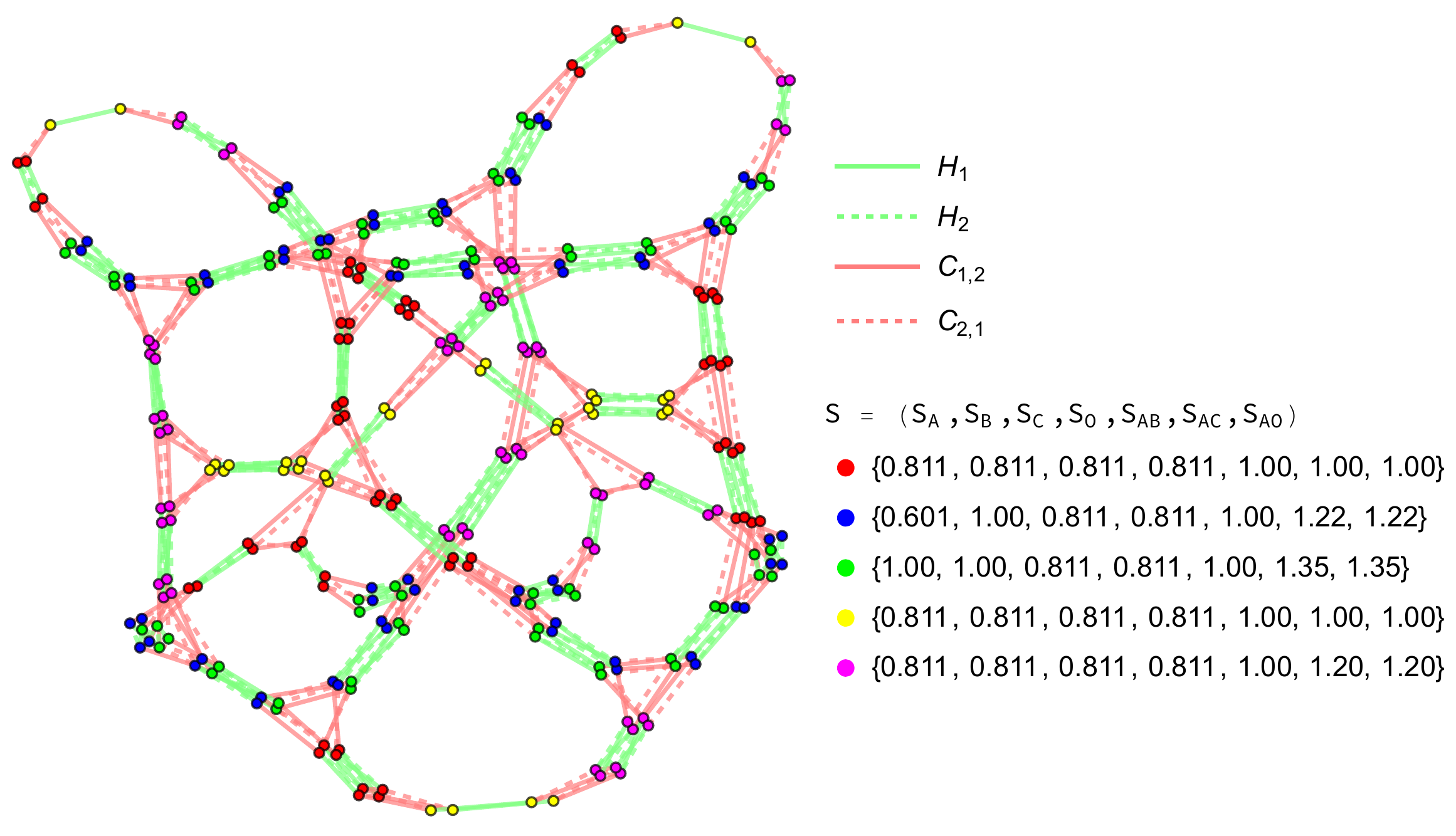}
        \caption{Orbit of $\ket{D^4_1}$, and stabilizer state, under the action of $(HC)_{1,2}$.}
    \label{HCGraphD41}
    \end{figure}

Below we include additional examples of $\ket{D^N_1}$ orbits under the action of $(HC)_{1,2} \equiv \langle H_1,\,H_2\,C_{1,2}\, C_{2,1}\rangle$. Figure \ref{HCGraphD41} shows the orbit for $\ket{D^4_1}$ under $(HC)_{1,2}$, which contains $288$ states and $4$ different entropy vectors. This set of $5$ entropy vectors is built of $6$ different entanglement entropies.

\begin{figure}[h]
        \centering
        \includegraphics[width=15cm]{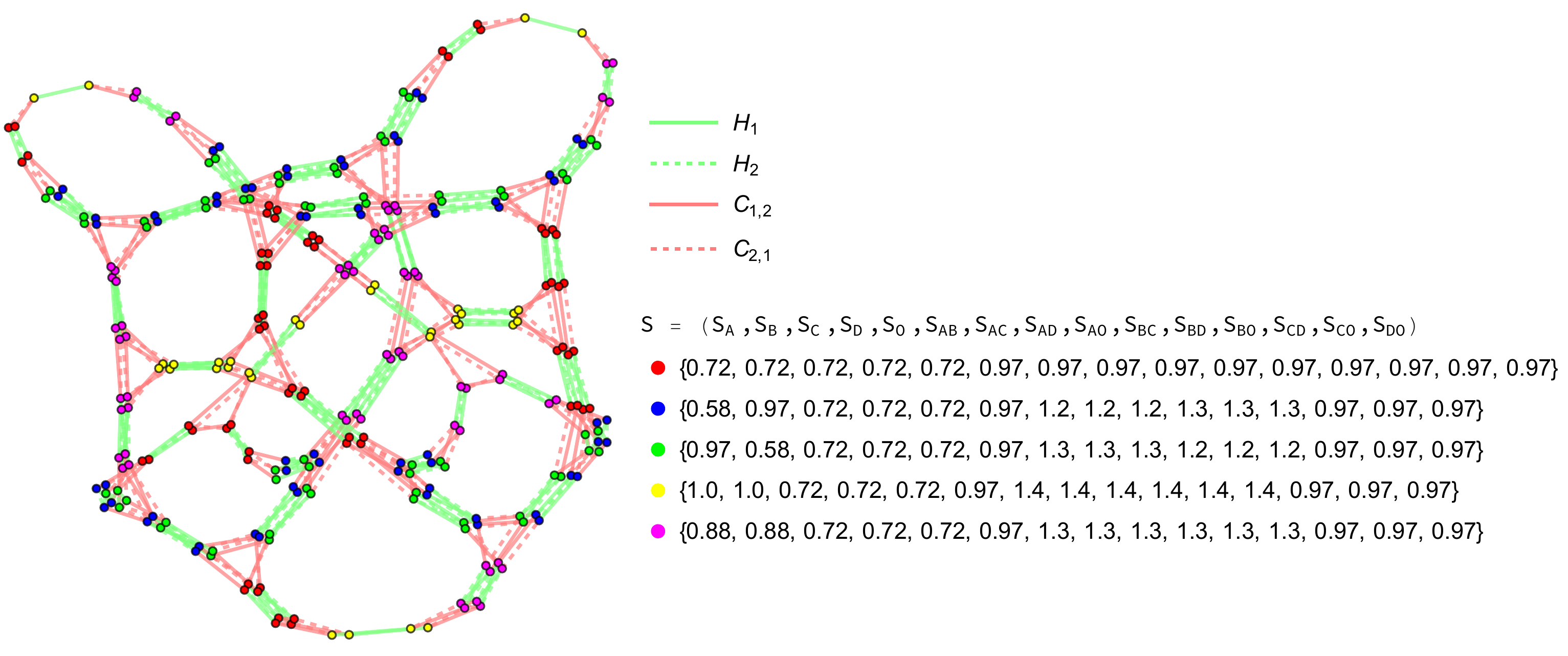}
        \caption{Orbit of $\ket{D^5_1}$, and stabilizer state, under the action of $(HC)_{1,2}$.}
    \label{HCGraphD51}
    \end{figure}

Figure \ref{HCGraphD51} gives the orbit of $\ket{D^5_1}$ under $(HC)_{1,2}$. This orbit likewise has $5$ different entropy vectors, which are composed of $9$ different entanglement entropies.

\section{Entropy Vector Tables}

In this Appendix we give exact entropy vectors seen in Figures \ref{HCGraphD31} and \ref{HCGraphD42}. Table \ref{tab:EntropyVectorTable1} gives each entropy vector from Figure \ref{HCGraphD31}. There are $4$ entanglement entropies observed in the orbit of $\ket{D^3_1}$ under the action of $\langle H_1,\,H_2\,C_{1,2}\, C_{2,1}\rangle$, which we define as variables in Eq.\ \eqref{Entropies31} for presentation clarity.
\begin{equation}\label{Entropies31}
    \begin{split}
        s_0&\equiv 1,\\
        s_1&\equiv \frac{2}{3}\log_2\left[\frac{3}{2}\right] +\frac{1}{3}\log_2\left[3\right],\\
        s_2&\equiv \frac{5}{6}\log_2\left[\frac{6}{5}\right] +\frac{1}{6}\log_2\left[6\right],\\
        s_3&\equiv \frac{3-\sqrt{5}}{6}\log_2\left[\frac{6}{3-\sqrt{5}}\right] +\frac{3+\sqrt{5}}{6}\log_2\left[\frac{6}{3+\sqrt{5}}\right],
    \end{split}
\end{equation}

\newpage

The four entropies in Eq.\ \eqref{Entropies31} build the entropy vectors in Table \ref{tab:EntropyVectorTable1}.
\begin{table}[h]
    \centering
    \begin{tabular}{|c||c|}
    \hline
    Label & Entropy Vector\\
    \hline
    \hline
    \fcolorbox{black}{red}{\rule{0pt}{6pt}\rule{6pt}{0pt}} & $(s_1,\,s_1,\,s_1)$\\
    \hline
    \fcolorbox{black}{blue}{\rule{0pt}{6pt}\rule{6pt}{0pt}} & $(s_3,\,s_1,\,s_1)$\\
    \hline
    \fcolorbox{black}{green}{\rule{0pt}{6pt}\rule{6pt}{0pt}} & $(s_1,\,s_3,\,s_1)$\\
    \hline
    \fcolorbox{black}{yellow}{\rule{0pt}{6pt}\rule{6pt}{0pt}} & $(s_0,\,s_0,\,s_1)$\\
    \hline
    \fcolorbox{black}{magenta}{\rule{0pt}{6pt}\rule{6pt}{0pt}} & $(s_2,\,s_2,\,s_1)$\\
    \hline
    \end{tabular}
\caption{The $5$ entropy vectors found in the $(HC)_{1,2}$ orbit of $\ket{D^3_1}$, shown in Figure \ref{HCGraphD31}. For brevity, we introduce the variables in Eq.\ \eqref{Entropies31} to present these entropy vectors.}
\label{tab:EntropyVectorTable1}
\end{table}

Similarly for the orbit of $\ket{D^4_2}$ under $\langle H_1,\,H_2\,C_{1,2}\, C_{2,1}\rangle$ action, there are $5$ entanglement entropies observe. We likewise define the variables,
\begin{equation}\label{Entropies42}
    \begin{split}
        s_0&\equiv \frac{5}{6}\log_2\left[\frac{12}{5}\right] +\frac{1}{6}\log_2\left[12\right],\\
        s_1&\equiv \frac{3-\sqrt{5}}{6}\log_2\left[\frac{12}{3-\sqrt{5}}\right] +\frac{3+\sqrt{5}}{6}\log_2\left[\frac{12}{3+\sqrt{5}}\right],\\
        s_2&\equiv \frac{2}{3}\log_2\left[\frac{3}{2}\right] +\frac{1}{3}\log_2\left[6\right],\\
        s_3&\equiv \frac{3-2\sqrt{2}}{6}\log_2\left[\frac{12}{3-2\sqrt{2}}\right] +\frac{3+2\sqrt{2}}{6}\log_2\left[\frac{12}{3+2\sqrt{2}}\right],\\
        s_4&\equiv 1,\\
        s_5&\equiv \frac{2}{3}\log_2\left[\frac{3}{2}\right] +\frac{1}{3}\log_2\left[3\right],\\
        s_6&\equiv \frac{5}{6}\log_2\left[\frac{6}{5}\right] +\frac{1}{6}\log_2\left[6\right].
    \end{split}
\end{equation}

\begin{table}[h]
    \centering
    \begin{tabular}{|c||c|}
    \hline
    Label & Entropy Vector\\
    \hline
    \hline
    \fcolorbox{black}{red}{\rule{0pt}{6pt}\rule{6pt}{0pt}} & $\left(s_4,\,s_4,\,s_4,\,s_4,\,s_2,\,s_2,\,s_2 \right)$\\
    \hline
    \fcolorbox{black}{blue}{\rule{0pt}{6pt}\rule{6pt}{0pt}} & $\left(s_6,\,s_5,\,s_4,\,s_4,\,s_2,\,s_1,\,s_1 \right)$\\
    \hline
    \fcolorbox{black}{green}{\rule{0pt}{6pt}\rule{6pt}{0pt}} & $\left(s_5,\,s_6,\,s_4,\,s_4,\,s_2,\,s_1,\,s_1 \right)$\\
    \hline
    \fcolorbox{black}{yellow}{\rule{0pt}{6pt}\rule{6pt}{0pt}} & $\left(s_4,\,s_4,\,s_4,\,s_4,\,s_2,\,s_0,\,s_0 \right)$\\
    \hline
    \fcolorbox{black}{magenta}{\rule{0pt}{6pt}\rule{6pt}{0pt}} & $\left(s_4,\,s_4,\,s_4,\,s_4,\,s_2,\,s_2,\,s_2 \right)$\\
    \hline
    \fcolorbox{black}{cyan}{\rule{0pt}{6pt}\rule{6pt}{0pt}} & $\left(s_6,\,s_6,\,s_4,\,s_4,\,s_2,\,s_3,\,s_3 \right)$\\
    \hline
    \end{tabular}
\caption{The $6$ entropy vectors contained in the orbit of $\ket{D^4_2}$ under the action of $(HC)_{1,2}$, illustrated in Figure \ref{HCGraphD42}. We introduce variables in Eq.\ \eqref{Entropies42} to display the entropy vectors in the table.}
\label{tab:EntropyVectorTable2}
\end{table}

\end{appendices}

\newpage

\bibliographystyle{JHEP}
\bibliography{DickeStates}

\end{document}